\documentclass[a4paper,11pt]{article}
\pdfoutput=1 

\usepackage{jheppub} 
\usepackage{bm}
\usepackage{dsfont}
\usepackage{slashed}
\usepackage[T1]{fontenc} 
\usepackage{graphicx}
\def\spa#1.#2{\left\langle#1\,#2\right\rangle}
\def\spb#1.#2{\left[#1\,#2\right]}

\def\cO{\mathcal{O}}
\def\cM{\mathcal{M}}

\def\cA{\mathcal{A}}
\def\cF{\mathcal{F}}

\def\qb{{\bar q}}
\def\ib{{\; \bar\imath}}

\def\e{\epsilon}
\def\tree{{\rm tree}}
\def\Soft{\mathop{\cal S}\nolimits}

\def\RS{{\scriptscriptstyle\rm R\!.S\!.}}
\def\bf#1{{\mbox{\boldmath $#1$}}}
\def\Li{\text{Li}}
\def\to{\rightarrow}
\def\slash#1{\ooalign{$\hfil/\hfil$\crcr$#1$}}
\def\bf#1{{\mbox{\boldmath $#1$}}}

\newcommand{\cV}       {{\cal V}}
\newcommand{\ra}{\rangle}

\newcommand{\Graph}[2]{\vcenter{\hbox{\includegraphics[scale=#1]{#2}}}}
\newcommand{\abs}[1]{\lvert#1\rvert}
\newskip\humongous \humongous=0pt plus 1000pt minus 100pt

\newif\ifdtup

\title{Double soft current at one-loop in QCD}

\author[a]{Yu~Jiao~Zhu}

\affiliation[a]{Zhejiang Institute of Modern Physics, Department of
  Physics, Zhejiang University, Hangzhou, 310027, China}
\emailAdd{ zhuyujiao@zju.edu.cn}

\abstract{
We investigate the soft behavior of QCD amplitudes involving multiple Wilson lines and derive compact analytic expressions for double soft gluon and double soft quark emissions at one loop. The color correlations of the soft current exhibit a predominantly dipole structure, coupling to two hard legs at a time, apart from an abelian contribution that factorizes into products of one-loop and tree-level single soft currents, which may involve up to three hard legs. The kinematic dependence of the one-loop soft currents is expressed in terms of polylogarithmic functions, with explicit results presented for time-like kinematics. We further discuss the analytic continuation to other kinematic configurations and identify non-trivial crossing effects when continuing into incoming states. The squared amplitude is found to be invariant under crossing, which implies that the fully differential soft function and in particular the TMD soft function, remains universal up to three loops.
}

\keywords{QCD amplitudes, Infrared factorization, Soft theorem}

\begin{document} 
\unitlength = 1mm
\maketitle
\flushbottom

\section{Introduction}
Multi-loop scattering amplitudes in gauge theories are intricate functions of the external momenta.
In certain kinematic limits, however, gauge-theory amplitudes exhibit enhanced symmetries that lead to remarkably simpler functional structures.
A particularly important situation arises when one or more external momenta become either soft or collinear.
In these limits, the amplitude factorizes into a universal emission factor and a lower-point hard amplitude.

In QCD, soft amplitudes capture the universal behavior of scattering processes in the limit where partons carry vanishing energy.
Their universality implies process independence, making them powerful tools for extracting information common to all QCD reactions.
Soft amplitudes constitute the essential amplitude-level building blocks of soft functions in effective field theories, which govern the factorization of soft-sensitive cross sections.
Prominent examples include the soft functions for the threshold production of Drell-Yan lepton pairs \cite{Belitsky:1998tc,Becher:2007ty},
Higgs bosons \cite{Duhr:2013msa,Li:2013lsa,Li:2014afw,Li:2014bfa,Anastasiou:2013mca,Anastasiou:2013srw},
top-quark pairs \cite{Ahrens:2010zv,vonManteuffel:2014mva,Wang:2018vgu},
and direct photons \cite{Becher:2009th},
as well as transverse-momentum-dependent (TMD) observables where soft recoils play a central role \cite{Li:2016ctv,Moult:2018jzp}.

Furthermore, the study of soft Wilson loops plays a central role for understanding the infrared (IR) structure of gauge theories \cite{Catani:1999ss,Bern:1998sc,Bern:1999ry,Becher:2009cu,Gardi:2009qi,Becher:2009qa,Dixon:2009ur,Almelid:2015jia,Almelid:2017qju,Falcioni:2019nxk}.
Through their renormalization, soft correlations determine the IR singularity structure of multi-leg amplitudes,
which in turn supplies the key ingredients for the resummation of large logarithms.
For precision QCD phenomenology, soft amplitudes also underpin the subtraction schemes that isolate infrared divergences at higher orders \cite{Catani:2007vq,Gehrmann-DeRidder:2005btv,Currie:2013vh,Czakon:2010td,Boughezal:2011jf,Boughezal:2015dva,Gaunt:2015pea,DelDuca:2016ily,Caola:2017dug,Bertolotti:2022aih},
and provide the universal kernels that govern parton-shower algorithms in Monte Carlo event generators \cite{Buckley:2011ms}.

In this work, we focus on the leading-power behavior of QCD amplitudes when two soft partons are emitted from an arbitrary number of hard partons.
At tree level, the double-soft amplitude can be decomposed into an Abelian contribution and an irreducible color-dipole correlation \cite{Catani:1999ss}.
We derive the one-loop quantum corrections to this tree-level double-soft amplitude, providing explicit results for both gluon and quark emissions.
Combined with existing results for the two-loop single-soft amplitude, which captures correlations among up to three hard partons \cite{Duhr:2013msa,Li:2013lsa,Li:2016ctv,Moult:2018jzp,Dixon:2019lnw},
and the tree-level triple-soft radiation \cite{Catani:2019nqv},
our results complete the picture of N$^3$LO soft correlations involving multiple hard lines at the amplitude level.
This provides the theoretical foundation for studying soft correlations in collider observables with multiple jets \cite{Gao:2019ojf}.

The remainder of this paper is organized as follows.
In Section \ref{sec:Tree}, we define the soft amplitude in terms of Wilson lines and review the known tree-level results.
Section \ref{sec:LOOP} presents explicit one-loop expressions for the double-soft gluon and quark amplitudes in color space.
In Section \ref{sec:SDE}, we introduce a novel approach for evaluating the relevant master integrals, and in Section \ref{sec:AC},
we perform the analytic continuation to  all physically relevant kinematic configurations.
\section{Soft Factorization and tree level amplitude\label{sec:Tree}}
A remarkable feature of gauge theories is that scattering amplitude involving a soft gauge
boson exhibit  universal factorization properties,
where a $n$-point amplitude with a  single soft emission factorizes into a color operator (see in~\ref{sec:colorSpace}) ${\bf J}^{a}$ acting on the $n-1$ point amplitude
without the soft gauge boson
\begin{align}
\label{eq:single soft fac}
\cM^{a}_n{{\buildrel q \to 0 \over\longrightarrow}}\,\, { \varepsilon}^{\mu}(q) {\bf J}^{a}_{\mu} \cdot \cM_{n-1}\,.
\end{align}
This is historically  known  as the soft theorem~\cite{Low:1958sn,Yennie:1961ad,Weinberg:1965nx}.
The one-loop radiative corrections to single soft emission  is  computed in~\cite{Catani:2000pi}, 
while the generalization to multiple soft emissions and to all loop orders was developed in Refs.~\cite{Bauer:2001yt,Feige:2014wja},
\begin{align}
\langle a_1,\dots,a_n|&\cM(q_1,\dots,q_n;p_1,\dots,p_m)\rangle|_{ \{q_1,\dots,q_n \} \to 0}
 \nonumber\\&
\,\longrightarrow\,\,{ \varepsilon}^{\mu_1}(q_1)\dots{ \varepsilon}^{\mu_n} (q_n)
{\bf J}^{a_1 \dots a_n}_{\mu_1\dots \mu_n}(q_1,\dots,q_n)|\cM(p_1,\dots,p_m)\rangle\,.
\label{frac theo}
\end{align}
The factorization formula holds at leading power in the soft limit, 
the explicit form of the color operator is\footnote{We cast the factorization formulae in a form as if the soft partons were gluons,
 for soft quarks one replace gluon polarizations and colors with those for the quarks.}
\begin{align}
{\varepsilon}^{\mu_1}(q_1)\dots{ \varepsilon}^{\mu_n}(q_n) {\bf J}^{a_1 \dots a_n}_{\mu_1 \dots \mu_n}(q_1,\dots,q_n)
&\equiv\langle q_1,\dots,q_n;a_1,\dots,a_n|\prod_{k=1}^n Y^{\dagger}_{n_k}(0) |\Omega\rangle\,,
\end{align}
where $n_k=p_k/p^0$ and $Y^{\dagger}_{n}$ corresponds to the outgoing soft Wilson line 
\begin{align}
Y^{\dagger}_{n} (x)\equiv P \left\{ \exp[  i g_s {\bf T}^a \int_{0}^{\infty}{n\cdot A^a(x + s n)e^{-\eta s}}{ds}]\right \}\,.
\label{eq:Wilson line}
\end{align}  
From  configuration where  all hard particles are outgoing one may consider analytic continuation into the configuration where part of hard particles are incoming, 
this is  achieved by replacing the outgoing quark/antiquark/gluon Wilson lines with the incoming ones.
In the Wilson line operator, we introduce   ${\bf T}^a$ as the color space generator~\cite{Catani:2000pi}, 
 and $\eta$ as the small  parameter for outgoing prescription.
Expanding the path-ordered exponential yields
\begin{align}
Y^{\dagger}_{n} 
=&\sum_{m=0}^\infty \frac{(i g_s)^m}{m!} \bf{T}_n^{a_m} \dots \bf{T}_n^{a_2} \bf{T}_n^{a_1} 
\int_0^\infty d s_1 \int_{s_1}^\infty d s_2 \dots \int_{s_{m-1}}^\infty d s_m e^{-\eta \sum_j s_j}
\cr\,&
 {\rm{T}}\left[ \,n\cdot A^{a_1}(s_1 n) n\cdot A^{a_2}(s_2 n) \dots n\cdot A^{a_m}(s_m n) \right]+
\text{permutations}\,.
\end{align}
Suppose  we have several Wilson lines evaluated at space-time point $x=0$,
 for simplicity we may consider two of them,
 the matrix element  of the product of the   Wilson lines between the vacuum and some arbitrary outgoing states  $\beta$ reads
\begin{align}
&\langle \beta| Y^{\dagger}_{n_i}Y^{\dagger}_{n_j}|\Omega\rangle
=\sum_{m=0}^\infty \frac{(i g_s)^m}{m!} \prod_{i=m}^1\bf{T}_{n_i}^{a_i} \sum_{l=0}^\infty \frac{(i g_s)^l}{l!} \prod_{j=l}^1\bf{T}_{n_j}^{b_j} 
\int d \vec{s}\int d \vec{t}\,\,e^{-\eta \sum_i s_i}e^{-\eta \sum_j t_j}
\cr\,&
\times\langle \beta|\,\, {\rm{T}}\,
\bigg[{n_i}\cdot A^{a_1}(s_1 {n_i}) \dots {n_i}\cdot A^{a_m}(s_m {n_i}) 
 \,\,n_j\cdot A^{b_1}(t_1 n_j) \dots n_j\cdot A^{b_l}(t_l n_j)  \bigg] |\Omega\rangle
 \cr\,
& +\text{permutations}
 \cr\,&
=\sum_{m=0}^\infty \frac{(i g_s)^m}{m!} \prod_{i=m}^1\bf{T}_{n_i}^{a_i}   \sum_{l=0}^\infty \frac{(i g_s)^l}{l!} \prod_{j=l}^1\bf{T}_{n_j}^{b_j}
\int_{k_i} \int_{q_j}\int d \vec{s}\,e^{-i (k_1\cdot {n_i}-i\eta) s_1}\dots e^{-i (k_m-i\eta)\cdot {n_i} s_m}
\cr\,&
\times\int d \vec{t} \,e^{-i (q_1\cdot n_j -i\eta)t_1}\dots e^{-i (q_l-i\eta)\cdot n_j t_l} 
\bm{\cF} \bigg[\langle\beta| {\rm T} \,n_i\cdot A^{a_1}(s_1 {n_i})  \dots  n_j\cdot A^{b_l}(t_l n_j)  |\Omega\rangle \bigg]
\cr\,
&+\text{permutations}
\cr\,&
=\sum_{{m,l}=0}^\infty \frac{(g_s)^{m+l}}{m! l!} \int_{k_i} \int_{q_j}\frac{\bf{T}_{n_i}^{a_m}\bf{T}_{n_i}^{a_{m-1}} \dots \bf{T}_{n_i}^{a_1} }
{{n_i}\cdot k_m {n_i}\cdot (k_m+k_{m-1})\dots {n_i}\cdot \sum k_i} \frac{\bf{T}_{n_j}^{b_l}\bf{T}_{n_j}^{b_{l-1}} \dots \bf{T}_{n_j}^{b_1} } 
{n_j\cdot q_l n_j\cdot (q_l+q_{l-1})\dots n_j\cdot \sum q_j}
\cr\,&
\times\bm{\cF} \bigg[\langle\beta| {\rm T} \,n_i\cdot A^{a_1}(s_1 {n_i})  \dots  n_j\cdot A^{b_l}(t_l n_j)  |\Omega\rangle \bigg]
+\text{permutations}\,,
\label{feynRule1}
\end{align}
where  we have introduced the shorthand notation
\begin{align}
\int d \vec{s}\equiv\int_0^\infty d s_1  \int_{s_1}^\infty d s_2 \dots \int_{s_{m-1}}^\infty d s_m\,,\quad\quad n_i\cdot (k_m+\dots)\to n_i\cdot (k_m+\dots)-i\eta\,.
\end{align}
In the second step of Eq.~(\ref{feynRule1}), we use the fact the fields along the directions  $n_i$ and $n_j$  are space-like separated, 
so  only  a single time-ordering operator $\rm T$ is required.
The symbol $\bm\cF$ denotes the Fourier transform of position-space Green function into momentum space.
The resulting expression in  the bracket then equals the sum of all Feynman diagrams with incoming state corresponding to vacuum $\Omega$, 
on-shell outgoing  lines  corresponding to states  in $\beta$, 
 and lines off the mass shell (including propagators) corresponding to the gauge field operators $ n_i\cdot A^{a_1} \dots n_j\cdot A^{b_l}$. 
 The multiplicity of the Green functions precisely cancels with the combinatoric  factors $m!\,l!$.
To illustrate the computation of the soft matrix element, we begin with a tree-level example of double soft gluons of momenta $q_{2,3}$~\cite{Catani:1999ss} ,
\begin{align}
\bf{J}_{a_2 a_3 }^{\mu_2\mu_3 (0)}(q_2,q_3)=  & \quad\Graph{1.}{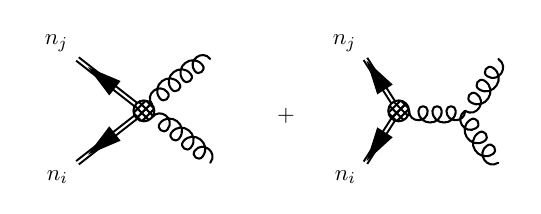}
\nonumber\\
=&
\frac{1}{2} \bigg\{ 
{\bf J}_{a_2}^{\mu_2 (0)}(q_2) \;, {\bf J}_{a_3}^{\mu_3 (0)}(q_3) \bigg\}
 + i f_{a_2a_3a} \sum_{i} {\bf T}_i^a \bigg\{
\frac{n_i^{\mu_2} q_2^{\mu_3} - n_i^{\mu_3} 
q_3^{\mu_2}}{(q_2\cdot q_3) \,[n_i\cdot (q_2+q_3)]}
 \nonumber\\&
- \frac{n_i\cdot (q_2-q_3)}{2 [n_i\cdot (q_2+q_3)]}
\left[ \frac{n_i^{\mu_2} n_i^{\mu_3}}{(n_i \cdot q_2) (n_i \cdot q_3)} 
+ \frac{g^{\mu_2\mu_3}}{q_2\cdot q_3} \right]\bigg \}\,,
\label{eq:TreeCurrent}
\end{align}
where  the bracket is the anticommutator of tree level single soft current 
\begin{align}
{\bf J}_{a}^{\mu (0)}(q)= - \sum_{i}
{\bf T}^{a}_i \frac{n_i^{\mu}}{n_i \cdot q}\,.
\label{eq:treesingle}
\end{align}
The current fulfills several properties:
\begin{itemize}
\item[a)] It is independent of the helicity and flavor of the massless hard partons, 
the only information it carries is the color charge and the direction of the hard scattered parton,
the latter property is dubbed ``rescaling invariance''.
\item[b)] Its divergence is proportional to the total color charge of the hard partons, 
which is a statement of on-shell gauge invariance
\footnote{The QED Ward Identity $q_\mu \cM^{\mu}(q) = 0$ does not require on-shell condition, but QCD does.}, see a detailed discussion in appendix~\ref{sec:colorSpace}
\begin{align}
q_{2\mu_2} {\bf J}_{a_2a_3 }^{\mu_2\mu_3(0)}(q_2,q_3) &=
\left(- {\bf J}_{a_3 }^{\mu_3 (0)}(q_3) \;\delta_{a_2 a} + \frac{i}{2}
f_{a_2a_3a} \frac{q_2^{\mu_3}}{q_2\cdot q_3} \right) 
\sum_{i=1}^{n} {\bf T}_i^a \;\;,
 \nonumber\\
q_{3\mu_3} {\bf J}_{a_2a_3 }^{\mu_2\mu_3 (0)}(q_2,q_3)& =
\left(- {\bf J}_{a_2 }^{\mu_2 (0)}(q_2) \;\delta_{a_3 a} + \frac{i}{2}
f_{a_3a_2a} \frac{q_3^{\mu_2}}{q_2\cdot q_3} \right) 
\sum_{i=1}^{n} {\bf T}_i^a \;\;.
\label{eq:gaugeInvar}
\end{align}
\end{itemize}
Using spinor helicity variables, the current can be converted into an amplitude. In color singlet basis for $\gamma^*\to q \bar q$,  it takes the following form
\begin{align}
\label{eq:doubletree}
\langle i_1\ib_4 |{\bf J}_{a_2a_3}^{\mu_2\mu_3 (0)}(q_2,q_3){\varepsilon_{\mu_2}}(q_2) {\varepsilon_{\mu_3}}(q_3)|\gamma^*\to q \qb\rangle \equiv 
\sum_{\sigma \in S_2}{ 2\,(t^{a_{\sigma(2)}}t^{a_{\sigma(3)}})^{~\ib_4}_{i_1} }\Soft^\tree(1_q,\sigma(2),\sigma(3),4_{\bar q})
\end{align} 
where  we label the external quark ($q$) and antiquark ($\qb$) as legs  1 and  4,  carrying fundamental color indices $i_1$ and $\ib_4$, 
respectively. The  tree-level form factors are
\begin{align}
\Soft^\tree(1,2^+,3^+,4) &= \frac{ \spa{1}.4}{ \spa{1}.2 \spa2.3 \spa3.4}\,,
 \nonumber\\
\Soft^\tree(1,2^+,3^-,4)& =\frac{1} {\spa{1}.2\spb2.4+ \spa1.3 \spb3.4\,}\left(\frac{1}{s_{12}
+s_{13} }\frac{\spb{1}.4 \spa1.3^3 }{ \spa{2}.3\spa1.2}+\frac{1}{ s_{24}+s_{34}} \frac{\spa{1}.4\spb2.4^3}{ \spb2.3 \spb3.4}\right)\,.
\label{eq:formf}
\end{align}
Other color and helicity components in Eq.~(\ref{eq:doubletree}) can be obtained by exchanging the external legs $1\leftrightarrow 4$ or by complex conjugation, 
for example
\begin{align}
\Soft^\tree(1,3^-,2^+,4)=\Soft^\tree(4,2^+,3^-,1)\,,\quad\quad \Soft^\tree(1,2^-,3^+,4)=\Soft^\tree(1,2^+,3^-,4)^*\,.
\end{align}
Similarly, the double soft quark-antiquark ($Q \bar Q$) emssion at tree level reads
\begin{align}
\label{eq:quark_tree}
({\bf J}^{(0)})_{i_2}^{~\ib_3}(q_2,q_3)\equiv\sum_{n}\frac{\slash{n}}{n\cdot (q_2+q_3)}\frac{1}{s_{23}}(t^{d})^{~\ib_3}_{i_2}\bf{T}_n^d \,,
\end{align}
where  $t^d$ is the color matrix in fundamental representation, and $(i_2,\ib_3)$ are color indices for quark $Q$ and  antiquark $\bar Q$, respectively. 
In  the color singlet basis  for $\gamma^*\to q \bar q$, we have
\begin{align}
\langle i_1\ib_4 |\bar u(q_3)({\bf J}^{(0)})_{i_2}^{~\ib_3}(q_2,q_3) v(q_2)|\gamma^*\to q \qb\rangle \equiv 
2\,(t^{d})^{~\ib_4}_{i_1} (t^{d})^{~\ib_3}_{i_2}\Soft^{\tree} (2\bar Q^+,3Q^-)\,,
\end{align}
\begin{align}
\Soft^{\tree} (2\bar Q^+,3Q^-)= \frac{1}{s_{23}}\Bigg(\frac{\spa{3}.1 \spb{1}.2}{s_{12}+s_{13}} - \frac{\spa{3}.4 \spb{4}.2}{s_{24}+s_{34}}\Bigg)\,.
\end{align}
\section{Double soft current at one-loop\label{sec:LOOP}}
\label{sec:double soft result}
 \begin{figure}[h]
  \begin{center}
    \includegraphics[width=\textwidth]{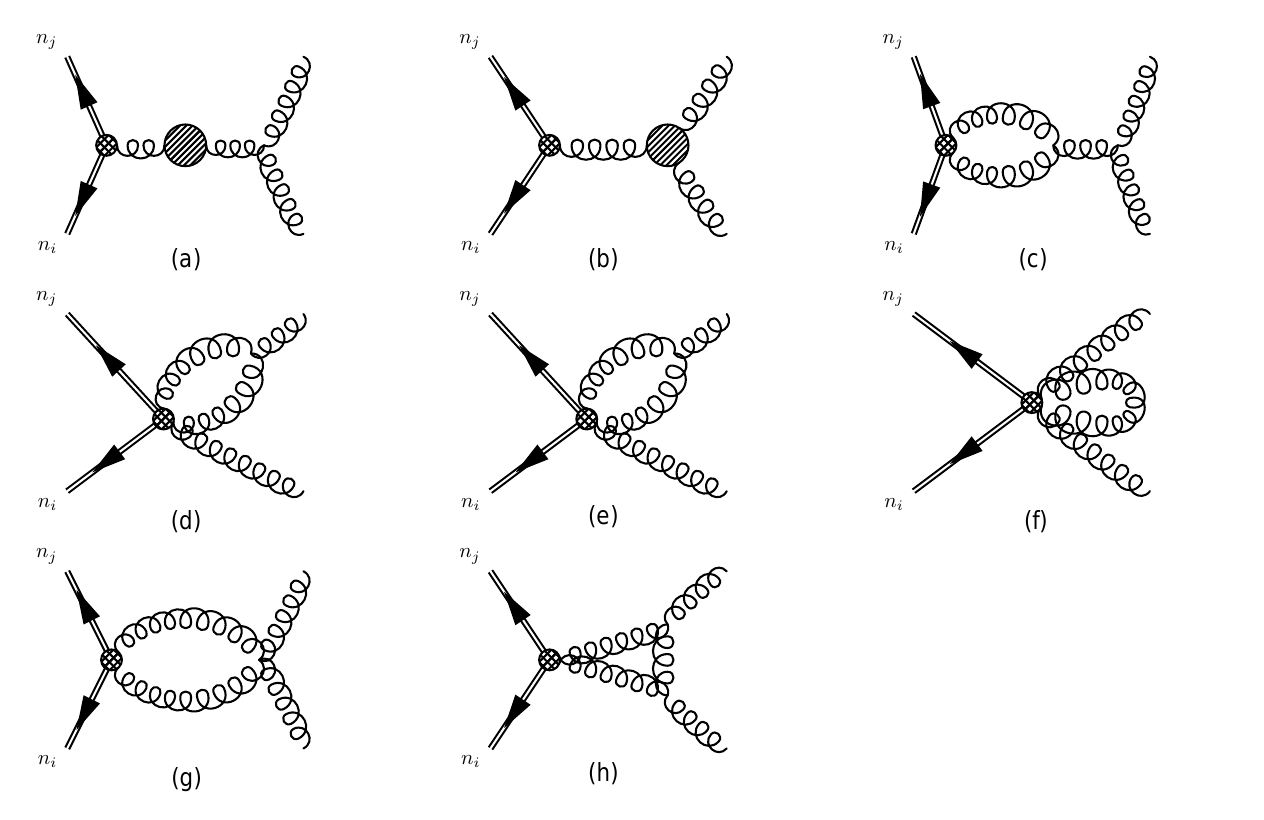}
     \includegraphics[width=\textwidth]{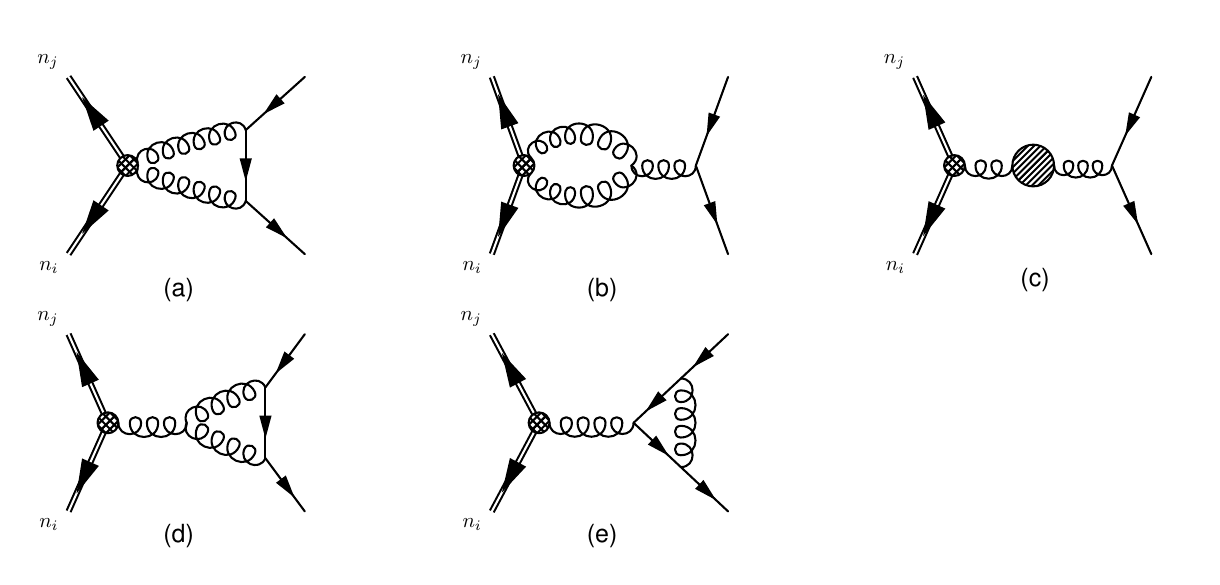}
  \end{center}
  \caption{Non-vanishing diagrams for double soft gluons and quarks in Feynman gauge.
   Diagram 4 and 5 can couple to three Wilson lines, but this contribution factorizes into the abelian part. }
  \label{fig:3}
\end{figure}
Beyond tree level, the soft amplitude receives quantum corrections. 
The single soft amplitude was first obtained in Ref.~\cite{Bern:1995ix} for color-ordered amplitudes,
and was later rederived in color space~\cite{Catani:2000pi} in axial gauge using  Catani--Grazzini soft insertion rules, see also~\cite{Bern:1999ry,Bern:1998sc,Dixon:2019lnw}. 
For double soft emissions, the presence of sub-leading color structures leads to  an  stronger entanglement of color and kinematics. 
Historically, analyses  in the 1990s  made exclusive use of \textquoteleft primitive amplitudes\textquoteright, 
which provide  a  starting point for  factorizing  one-loop amplitudes in the soft and collinear regions by disentangling color  from  kinematics~\cite{Bern:1998sc,Bern:1999ry}. 
A representative example of   such a decomposition  appears in the artwork of helicity amplitudes for $\gamma*\to4$ partons~\cite{Bern:1997sc,Bern:1996ka}.
From there one can in principle, extract the explicit results for double soft emissions  (as we have done), 
however,  the procedure is tedious due to  momentum-conservation constraints in the helicity representation and the nonlinear   Schouten identities.
This motivates a direct calculation in color space, which we will  present below.   
\subsection{Setups}  
 Expanding both sides of Eq.~(\ref{frac theo}) to one-loop order in terms of bare strong coupling for double soft emission, 
 we obtain our starting formula:
\begin{align}
&\langle a_2,a_3|\cM^{(1)}{(q_2,q_3;p_1,\dots,p_m)}\rangle|_{ q_2 \to 0,q_3 \to 0} 
={ \varepsilon}^{\mu_2}(q_2){ \varepsilon}^{\mu_3} (q_3)g_s^2
\nonumber\\&
\times\left[{\bf J}^{a_2 a_3 (0)}_{\mu_2 \mu_3}(q_2,q_3)|\cM^{(1)}{(p_1,\dots,p_m)}\rangle
+\bar a \,{\bf J}^{a_2 a_3 (1)}_{\mu_2 \mu_3}(q_2,q_3)|\cM^{(0)}{(p_1,\dots,p_m)}\rangle\right]\,,
\label{eq:start}
\end{align}
where  the tree level current is given in Eq.~(\ref{eq:TreeCurrent}).
We  factor out a tree-level normalization of $g_s^2$ from the soft current, and expand  in the rescaled coupling 
\begin{align}
\bar a\equiv\frac{g_s^2 \mu^{-2\epsilon}}{(4 \pi)^{2-\e}}e^{-\e \gamma_E}=\frac{\alpha_s}{4 \pi}\frac{ \mu^{-2\epsilon} e^{-\e \gamma_E}}{(4 \pi)^{-\e}}\,,
\end{align}
where $\alpha_s=g_s^2/(4 \pi)$, and $\gamma_E=0.577216$ is the Euler-Mascheroni constant.

We generate Feynman diagrams in Fig.~(\ref{fig:3})  by \texttt{Qgraph}~\cite{Nogueira:1991ex}. 
The color/Dirac algebra and integrand manipulations are performed in   \texttt{form}~\cite{Ruijl:2017dtg}. 
To  track  regularization-scheme dependence we set regularization-dependent dimension to  $4-2\delta_R \epsilon$, 
with $\delta_R=0$ defining the four dimensional helicity scheme ($\rm FDH $)~\cite{Bern:1991aq,Bern:2002zk} and $\delta_R=1$ defining 't~Hooft-Veltman scheme ($\rm HV$)~\cite{tHooft:1972tcz}.
 The $d$-dimension loop reduction is based on IBP identities~\cite{Chetyrkin:1981qh} implemented in   \texttt{LiteRed}~\cite{Lee:2012cn}.   
 Below we present   the soft amplitude with a normalization 
\begin{align}
c_\Gamma\equiv{\Gamma(1+\epsilon)\Gamma^2(1-\epsilon)\over \Gamma(1-2\epsilon)}\,.
\end{align}
We    assume time-like kinematics ($s_{i j}>0$) and later discuss analytic continuation.
\subsection{Time-Like results for double soft gluons}
For double soft gluons, we decompose the amplitudes into two independent helicity configurations (others followed by complex conjugation).
Analogous to the   tree-level case  Eq.~(\ref{eq:TreeCurrent}), 
 the amplitude   splits into an abelian part,
which is a direct product of a one-loop and tree-level single soft emission,
  and a correlated non-abelian  part 
\begin{align}
{ \varepsilon}_{\mu_2}(q_2){ \varepsilon}_{\mu_3}(q_3)\bf{J}_{a_2 a_3 }^{\mu_2 \mu_3 (1)}(q_2,q_3)=\cM_{a_2 a_3}^{\text{ab.}}(2,3)+\cM_{a_2 a_3}^{\rm{nab.}}(2,3).
\end{align}
\begin{align}
\Delta^{\mu}(i,j;q;\e)\equiv&{{  \Gamma(1-\epsilon)\Gamma(1+\epsilon)}\over{\epsilon}^2}\Bigg({\mu^2(-s_{i j}-i\eta)\over{(-s_{i q}-i\eta)\,\,(-s_{q j} -i\eta)\,}}\Bigg)^{\epsilon}
\left(\frac{n_i^{\mu}}{n_i \cdot q}-\frac{n_j^{\mu}}{n_j \cdot q}\right)\,,
\cr\,
\cM_{a_2 a_3}^{\rm{ab.}}(2,3)=&
\sum_{i \neq j}^m i f_{a_2 c d}\bf{T}^c_{i}\bf{T}^d_{j}\Delta^{\mu_2}(i,j;q_2;\e){ \varepsilon}_{\mu_2}(q_2)
\sum_k^m \frac{-n_k\cdot{ \varepsilon}(q_3)}{n_k\cdot q_3} \bf{T}_k^{a_3}
 + 2 \longleftrightarrow 3 \,.
 \label{eq:ab.}
\end{align}
\begin{align}
&\cM_{a_2 a_3}^{\rm{nab.}}(2^+ ,3^+)=
-4 { \left(\frac{-s_{2 3}-i\eta}{\mu^2}\right)^{-\epsilon}}\sum_{i \neq j}^m  f_{b c a_2}f_{b d a_3} \bf{T}^c_{i}\bf{T}^d_{j} {\spa i.j \over \spa i.2 \spa 2.3 \spa3.j}
\cr
& \times\Bigg\{ {-2\over \epsilon^2} - {1\over \epsilon} \ln{s_{ij} s_{2 3}\over s_{i2} s_{3j}}
- {1\over 2}\Bigg(\ln^2{s_{ij}s_{2 3} \over (s_{i2}+s_{i3})(s_{2j}+s_{3j})}+\ln^2{ s_{i2}\over s_{i2} + s_{i3}}+\ln^2{s_{3j}\over s_{2j}+s_{3j}}\Bigg )\,
\cr
& - \Li_2(1- { s_{i2}\over s_{i2} + s_{i3}})-\Li_2(1- {s_{3j}\over s_{2j}+s_{3j}})-\Li_2(1- {s_{ij}s_{2 3} \over (s_{i2}+s_{i3})(s_{2j}+s_{3j})})
\Bigg\}
\cr
&-{ \left(\frac{-s_{2 3}-i\eta}{\mu^2}\right)^{-\epsilon}}
\Bigg\{{ \frac{C_A-n_f-C_A \delta_R \epsilon}{(-1+\epsilon)(-3+2\epsilon)(-1+2\epsilon)} } \sum_{i}^m  i f_{b a_2 a_3} \bf{T}_i^b\, {\frac{s_{i2}-s_{i3}}{s_{i2}+s_{i3}}}\frac{-1}{\spa 2.3^2}
\Bigg\} \,,
\label{eq:plusplus}
\end{align}
\begin{align}
&\cM_{a_2 a_3}^{\rm{nab.}}(2^+ ,3^-)=
-4 { \left(\frac{-s_{2 3}-i\eta}{\mu^2}\right)^{-\epsilon}}\sum_{i \neq j}^m  f_{b c a_2}f_{b d a_3} \bf{T}^c_{i}\bf{T}^d_{j} 
\cr
 &\times\Bigg\{{ {1 \over \spa{i}.2\spb2.j+ \spa i.3 \spb3.j\,}\left({1\over s_{i2}+s_{i3} }{\spb{i}.j \spa i.3^3 \over \spa{2}.3\spa i.2}+{1\over s_{2j}+s_{3j}}{\spa{i}.j\spb2.j^3 \over \spb2.3 \spb3.j}\right)} 
 \cr
&\times\Bigg({-2\over \epsilon^2} - {1\over \epsilon} \ln{s_{ij} s_{2 3}\over s_{i2} s_{3j}}
+{1\over 2} \ln^2{s_{ij} s_{2 3}\over s_{i2} s_{3j}} - {\pi^2\over 3}\Bigg) 
\cr
&+{\spa i.j\over \spa i.2 \spa2.j }{\spb i.j\over \spb i.3 \spb3.j }
\times\left({1\over2} \ln^2{s_{ij} s_{2 3}\over s_{i2} s_{3j}} \right)
\cr
&+ { 1\over s_{i2}+s_{i3}}\Bigg({  -1 \over \spa{i}.2\spb2.j+ \spa i.3 \spb3.j\,}  { \spb{i}.j \spa i.3^3 \over \spa{2}.3\spa i.2}   + {  -1 \over \spb{i}.2\spa2.j+ \spb i.3 \spa3.j\,}  { \spa{i}.j \spb i.2^3 \over \spb{2}.3\spb i.3}  \Bigg) 
\cr
&\times\Bigg({1\over 2} \ln^2{s_{i2}\over s_{i2} +s_{i3}  }+ {1\over 2} \ln^2{s_{ij} s_{2 3}\over (s_{i2} +s_{i3} )s_{3j} }\Bigg) 
\cr
&+{ 1\over s_{2j}+s_{3j}}\Bigg({  -1 \over \spa{i}.2\spb2.j+ \spa i.3 \spb3.j\,}  { \spa{i}.j \spb2.j^3 \over \spb{2}.3\spb3.j} 
+{  -1 \over \spb{i}.2\spa2.j+ \spb i.3 \spa3.j\,} { \spb{i}.j \spa3.j^3 \over \spa{2}.3\spa2.j} \Bigg) 
\cr
&\times\Bigg({1\over 2} \ln^2{s_{3j}\over s_{2j} +s_{3j}  } + {1\over 2} \ln^2{s_{ij} s_{2 3}\over (s_{2j} +s_{3j} )s_{i2} }\,\Bigg)\Bigg\}\,,
\label{eq:plusminus}
\end{align} 
\begin{align} 
\label{eq:strong-order}
&\cM_{a_2 a_3}^{\rm{nab.}}(2^+ ,3^+,2 \ll 3) =4  \sum_{i \neq j}^m  f_{b c a_2}f_{b d a_3} \bf{T}^c_{i}\bf{T}^d_{j} {\spa i.3 \over \spa i.2 \spa2.3} {\spa i.j \over \spa i.3 \spa3.j}\,
\cr
& \Bigg[\Bigg( {1\over \epsilon^2 } + {1\over \epsilon} \ln{-s_{i3}\,\mu^2\over s_{i2}\,s_{2 3}} +{1\over 2} \ln^2{-s_{i3}\,\mu^2\over s_{i2}\,s_{2 3}} + {\pi^2\over 6}\Bigg)
+\Bigg( {1\over \epsilon^2 } + {1\over \epsilon} \ln{-s_{ij}\,\mu^2\over s_{i3}\,s_{3j}} + {1\over 2} \ln^2{-s_{ij}\,\mu^2\over s_{i3}\,s_{3j}} + {\pi^2\over 6}\Bigg)\Bigg]\,,
\cr\,
&\cM_{a_2 a_3}^{\rm{nab.}}(2^+ ,3^-,2 \ll 3) =4  \sum_{i \neq j}^m  f_{b c a_2}f_{b d a_3} \bf{T}^c_{i}\bf{T}^d_{j} {\spa i.3 \over \spa i.2 \spa2.3} {-\spb i.j \over \spb i.3 \spb3.j}\,
\cr
& \Bigg[\Bigg( {1\over \epsilon^2 } + {1\over \epsilon} \ln{-s_{i3}\,\mu^2\over s_{i2}\,s_{2 3}} +{1\over 2} \ln^2{-s_{i3}\,\mu^2\over s_{i2}\,s_{2 3}} + {\pi^2\over 6}\Bigg)
+\Bigg( {1\over \epsilon^2 } + {1\over \epsilon} \ln{-s_{ij}\,\mu^2\over s_{i3}\,s_{3j}} + {1\over 2} \ln^2{-s_{ij}\,\mu^2\over s_{i3}\,s_{3j}} + {\pi^2\over 6}\Bigg)\Bigg]\,.
\cr\,
\end{align}
A particular case  is  the strongly ordered limit of the emitted soft gluons.
Taking   $q_2^0\ll q_3^0$, 
the  strongly ordered double soft gluon current is obtained by successive application of the single soft factorization formula Eq.~(\ref{eq:single soft fac}) :
\begin{align}
\langle a_2 \,a_3 | \cM \rangle\to \langle a_3 |  \bf{\tilde J}^{a_2}_{\mu_2}|\cM\rangle= \langle a_3 |  \bf{\tilde J}^{a_2}_{\mu_2}|a\rangle\langle a|\cM\rangle
\to \langle a_3 |  \bf{\tilde J}^{a_2}_{\mu_2}|a\rangle \bf{ J}^{a}_{\mu_3} |\cM\rangle\,,
\end{align}
hence
\begin{align}
\label{eq:strong-ordered}
\bf{J}_{a_2 a_3 }^{\mu_2 \mu_3 }(q_2\ll q_3)=
\langle a_3 |  \bf{\tilde J}^{a_2}_{\mu_2}|a\rangle \bf{ J}^{a}_{\mu_3}\,.
\end{align}
Here $\bf{\tilde J}^{a_2}_{\mu_2}$ acts on the color space of \text{`soft gluon $3$ + hard partons'}.
And the current $\langle a_3 |  \bf{\tilde J}^{a_2}_{\mu_2}|a\rangle$ acting on the color space of the hard partons,  
is partially contracted in the color indices of  soft gluon $3 $, so that its multiplication with the current $\bf{ J}^{a}_{\mu_3}$ make sense,
it is written as 
\begin{align}
\langle a_3 |  \bf{\tilde J}^{a_2}_{\mu_2}|a\rangle= 
\bf{ J}^{ a_2}_{\mu_2}\langle a_3  | a\rangle+\Delta \bf{ J}^{a;a_2,a_3}_{\mu_2}
=\bf{ J}^{ a_2}_{\mu_2}\delta_{a_3 a}+\Delta \bf{ J}^{a;a_2,a_3}_{\mu_2}\,,
\end{align}
where the current $\bf{ J}^{ a_2}_{\mu_2}$ is the standard single soft current and $\Delta \bf{ J}^{a;a_2,a_3}_{\mu_2}$ encodes nontrivial   color  correlations with soft gluon $3$.
At tree-level~\cite{Catani:1999ss},
\begin{align}
\bf{J}_{a_2 a_3 }^{ (0) \mu_2 \mu_3 }(q_2\ll q_3)=&\left(\bf{ J}^{(0) a_2}_{\mu_2}\langle a_3  | a\rangle+\langle a_3  |T_3^{a_2}| a\rangle \frac{q_{3 \mu_2}}{-q_3 \cdot q_2}\right) \bf{J}^{(0) a}_{\mu_3}
\cr\,
 =&\left(\bf{ J}^{(0) a_2}_{\mu_2}\delta_{a_3 a}+i f_{a_3 a_2 a} \frac{q_{3 \mu_2}}{-q_3 \cdot q_2}\right) \bf{J}^{(0) a}_{\mu_3}\,,
 \cr\,
\Delta \bf{ J}^{(0) a;a_2,a_3}_{\mu_2}=&i f_{a_3 a_2 a} \frac{q_{3 \mu_2}}{-q_3 \cdot q_2}\,.
\label{eq:dtree}
\end{align}
 At one-loop, expanding Eq.~({\ref{eq:strong-ordered}}) to one-loop order we obtain
 \begin{align}
 \bf{J}^{(1) a_2 a_3 }_{  \mu_2 \mu_3 }(q_2\ll q_3)
 =& \langle a_3| \bf{\tilde J}^{(0) a_2}_{\mu_2} |a\rangle \bf{J}^{(1) a}_{\mu_3}+
 \langle a_3| \bf{\tilde J}^{(1) a_2}_{\mu_2} |a\rangle \bf{J}^{(0) a}_{\mu_3}
 \cr\,
 =&\left[
  \bf{J}^{(0) a_2}_{\mu_2} \delta_{a_3 a}+\Delta \bf{ J}^{(0) a;a_2,a_3}_{\mu_2}
 \right]  \bf{J}^{(1) a}_{\mu_3}
 +
 \left[
  \bf{J}^{(1) a_2}_{\mu_2} \delta_{a_3 a}+\Delta \bf{ J}^{(1) a;a_2,a_3}_{\mu_2}
 \right]  \bf{J}^{(0) a}_{\mu_3}
 \cr\,
 =&\bf{J}^{(0)a_2}_{\mu_2}\bf{J}^{(1)a_3}_{\mu_3}+
 \bf{J}^{(1)a_2}_{\mu_2}\bf{J}^{(0)a_3}_{\mu_3}+
 \Delta \bf{ J}^{(0) a;a_2,a_3}_{\mu_2} \bf{J}^{(1) a}_{\mu_3}+
 \Delta \bf{ J}^{(1) a;a_2,a_3}_{\mu_2}\bf{J}^{(0) a}_{\mu_3}
 \cr\,
 =&\bigg\{\bf{J}^{(1)a_3}_{\mu_3}\bf{J}^{(0)a_2}_{\mu_2}+
 \bf{J}^{(1)a_2}_{\mu_2}\bf{J}^{(0)a_3}_{\mu_3}\bigg\}
 +
\bigg\{ \left[\bf{J}^{(0)a_2}_{\mu_2},\bf{J}^{(1)a_3}_{\mu_3}\right]
 +\Delta \bf{ J}^{(0) a;a_2,a_3}_{\mu_2} \bf{J}^{(1) a}_{\mu_3}\bigg\}
   \cr\,+&
 \Delta \bf{ J}^{(1) a;a_2,a_3}_{\mu_2}\bf{J}^{(0) a}_{\mu_3}\,.
 \end{align}
The first curly  bracket is the abelian contribution.
The remaining contributions are intrinsically non-abelian, and are individually gauge invariant:
\begin{align}
q_2^{\mu_2}\Delta \bf{ J}^{(1) a;a_2,a_3}_{\mu_2}\bf{J}^{(0) a}_{\mu_3}\simeq  0 \,,\quad\quad q_2^{\mu_2}\left(\left[\bf{J}^{(0)a_2}_{\mu_2},\bf{J}^{(1)a_3}_{\mu_3}\right]+\Delta \bf{ J}^{(0) a;a_2,a_3}_{\mu_2} \bf{J}^{(1) a}_{\mu_3}\right)\simeq 0\,.
\end{align} 
 The last non-abelian term  reads
 \begin{align}
\Delta \bf{ J}^{(1) a;a_2,a_3}_{\mu_2} \bf{J}^{(0) a}_{\mu_3}=&
2\sum_{i=1}^m i f_{a_2 c d } \langle a_3| \bf {T}_3^c |a \rangle  \bf {T}_i^d \Delta_{\mu_2}(3,i;q_2;\e)
\sum_{j=1}^m \frac{-{n_j}_{\mu_3}}{n_j\cdot q_3} \bf{T}_j^{a}
\cr\,
=&-2\sum_{i=1}^m f_{a_2 c d } f_{a_3 c a} \bf {T}_i^d \Delta_{\mu_2}(3,i;q_2;\e)
\sum_{j=1}^m \left(\frac{{n_i}_{\mu_3}}{n_i\cdot q_3}-\frac{{n_j}_{\mu_3}}{n_j\cdot q_3}\right) \bf{T}_j^{a}
\cr\,
=&2\sum_{i\neq j}^m f_{b c a_2} f_{b d a_3}  \bf {T}_i^c  \bf {T}_j^d \Delta_{\mu_2}(i,3;q_2;\e) \left(\frac{{n_i}_{\mu_3}}{n_i\cdot q_3}-\frac{{n_j}_{\mu_3}}{n_j\cdot q_3}\right)\,.
  \label{eq:{1contri}}
 \end{align}
 The   commutator piece of the   non-abelian term gives 
 \begin{align}
  \left[\bf{J}^{(0)a_2}_{\mu_2},\bf{J}^{(1)a_3}_{\mu_3}\right]
  = &\sum_k^m\sum_{i\neq j}^m 
 \frac{-{n_k}_{\mu_2}}{n_k\cdot q_2} i f_{a_3 c d} \Delta_{\mu_3}(i,j;q_3;\e) 
  \left[
  \bf{T}_k^{a_2},
   \bf{T}_i^{c} \bf{T}_j^{d}
  \right]
  \cr\,
  =&\sum_k^m\sum_{i\neq j}^m  \frac{-{n_k}_{\mu_2}}{n_k\cdot q_2} i f_{a_3 c d} \Delta_{\mu_3}(i,j;q_3;\e) 
  \times
  \left(
  i f_{a_2 c e} \bf{T}_k^e \bf{T}_j^d \delta_{ki}+
   i f_{a_2 d e} \bf{T}_i^c \bf{T}_k^e \delta_{kj}
  \right)
  \cr\,
  =&-2\sum_{i\neq j}^m 
   f_{b c a_2} f_{b d a_3}  \bf{T}_i^c \bf{T}_j^d
  \Delta_{\mu_3}(i,j;q_3;\e)  \frac{-{n_i}_{\mu_2}}{n_i\cdot q_2}\,.
 \end{align}
The  first  non-abelian term gives an overall contribution as follows 
\begin{align}
\left[\bf{J}^{(0)a_2}_{\mu_2},\bf{J}^{(1)a_3}_{\mu_3}\right]+\Delta \bf{ J}^{(0) a;a_2,a_3}_{\mu_2} \bf{J}^{(1) a}_{\mu_3}
=&
-2\sum_{i\neq j}^m 
   f_{b c a_2} f_{b d a_3}  \bf{T}_i^c \bf{T}_j^d
  \Delta_{\mu_3}(i,j;q_3;\e)  \frac{-{n_i}_{\mu_2}}{n_i\cdot q_2}
  \cr\,
  +&
  \sum_{i\neq j}^m  f_{a_3 a_2 a} f_{a c d} \bf{T}_i^c \bf{T}_j^d  \Delta_{\mu_3}(i,j;q_3;\e) \frac{{q_3}_{\mu_2}}{q_3\cdot q_2}
  \cr\,
  =&
  2\sum_{i\neq j}^m 
   f_{b c a_2} f_{b d a_3}  \bf{T}_i^c \bf{T}_j^d
  \Delta_{\mu_3}(i,j;q_3;\e)  \left(\frac{{n_i}_{\mu_2}}{n_i\cdot q_2}- \frac{{q_3}_{\mu_2}}{q_3\cdot q_2}\right)\,,
  \cr
  \label{eq:{2contri}}
\end{align}
where in the last step we have used the identity 
\begin{align}
\sum_{i\neq j}^m ( f_{b c a_2} f_{b d a_3}+ f_{b c a_3} f_{b d a_2}) \bf{T}_i^c \bf{T}_j^d \Delta_{\mu_3}(i,j;q_3;\e) = 0\,.
\end{align}
Adding Eq.~(\ref{eq:{1contri}}) and Eq.~(\ref{eq:{2contri}}) yields 
\begin{align}
\bf{J}^{(1) a_2 a_3 }_{  \mu_2 \mu_3 }(q_2&\ll q_3)|_{\rm{nab.}}=
 2\sum_{i\neq j}^m   f_{b c a_2} f_{b d a_3}  \bf{T}_i^c \bf{T}_j^d
  \left(\frac{{n_i}_{\mu_2}}{n_i\cdot q_2}- \frac{{q_3}_{\mu_2}}{q_3\cdot q_2}\right)
  \times
  \left(\frac{{n_i}_{\mu_3}}{n_i\cdot q_3}-\frac{{n_j}_{\mu_3}}{n_j\cdot q_3}\right)
  \cr\,
  \times&
  {{  \Gamma(1-\epsilon)\Gamma(1+\epsilon)}\over{\epsilon}^2}
  \left[
  \Bigg({\mu^2(-s_{i 3}-i\eta)\over{(-s_{i 2}-i\eta)\,\,(-s_{2 3 } -i\eta)\,}}\Bigg)^{\epsilon}
  +
  \Bigg({\mu^2(-s_{i j}-i\eta)\over{(-s_{i 3}-i\eta)\,\,(-s_{3 j} -i\eta)\,}}\Bigg)^{\epsilon}
  \right]\,,
\end{align}
 in full agreement with Eq.~(\ref{eq:strong-order}) from a direct computation of the   strongly ordered limit of the full results
\footnote{${n_i \cdot \varepsilon^+(q)}/{n_i \cdot q}-{n_j \cdot \varepsilon^+(q)}/{n_j \cdot q}=-\sqrt 2 {\langle i j \rangle}/ ( {\langle i q \rangle \langle q j \rangle  )}$}.
 
\subsection{Time-Like results for double soft quarks}
The soft quark amplitude is decomposed into four gauge invariant building blocks, 
a piece of uniform transcendental weight $\cM^{\rm {u.t.}}$, a piece which violates uniform transcendentality $\cM^{\rm {u.t.v.}}$, 
a subleading-color contribution given by the last triangle diagram in Fig.~(\ref{fig:3}) $\cM^{\rm {s.l.}}$, and the $n_f$ term $\cM^{\rm {nf}}$. 
Amplitudes for the helicity configuration $\bar Q^-Q^+$ can be obtained by complex conjugation\footnote{Colors left untouched, 
the complex conjugation is  implemented only on the kinematical part. }.
\begin{align}
\bar u(q_3)({\bf J}^{(1)})_{i_2}^{~\ib_3}(q_2,q_3) v(q_2)=
(\cM^{\rm {u.t.}})_{i_2}^{~\ib_3}+(\cM^{\rm{u.t.v.}})_{i_2}^{~\ib_3}
+(\cM^{\rm{s.l.}})_{i_2}^{~\ib_3}+(\cM^{\rm{n_f}})_{i_2}^{~\ib_3}\,.
\end{align}
\begin{align}
\label{eq:softQuark}
&(\cM^{\rm {u.t.}})_{i_2}^{~\ib_3}(2^+,3^-)=\cr\,
&4{ \left(\frac{-s_{2 3}-i\eta}{\mu^2}\right)^{-\epsilon}}\sum_{i\neq j}^m ( t^d t^c)^{~\ib_3}_{i_2} \bf{T}_{i}^c \bf{T}_{j}^d 
\Bigg\{\frac{-1}{s_{23}}\Bigg(\frac{\spa{3}.i \spb{i}.2}{s_{i2}+s_{i3}} - \frac{\spa{3}.j \spb{j}.2}{s_{2j}+s_{3j}}\Bigg)
\Bigg(- \frac{1}{\epsilon^2} - \frac{1}{\epsilon}\ln \frac{s_{ij} s_{23}}{s_{i2} s_{3j}}\Bigg)
\cr
&+\Bigg(\frac{\spa{i}.j \spb{i}.2^2}{(s_{i2}+s_{i3})(\spb{i}.2 \spa{2}.j + \spb{i}.3 \spa{3}.j)\spb{2}.3}-\frac{\spb{i}.j \spa{i}.3^2}{(s_{i2}+s_{i3})(\spa{i}.2 \spb{2}.j + \spa{i}.3 \spb{3}.j)\spa{2}.3}\Bigg)
\cr
&\times\Bigg(\frac{\pi^2}{6} + \frac{1}{2} \ln^2\frac{s_{i2}}{s_{i2}+s_{i3}} + \frac{1}{2} \ln^2\frac{s_{ij}s_{23}}{(s_{i2}+s_{i3})s_{3j}} - \frac{1}{4} \ln^2 \frac{s_{ij} s_{23}}{s_{i2}s_{3j}} \Bigg)
\cr
&+\Bigg(\frac{\spa{i}.j \spb{2}.j^2}{(s_{2j}+s_{3j})(\spa{i}.2 \spb{2}.j + \spa{i}.3 \spb{3}.j)\spb{2}.3}-
\frac{\spb{i}.j \spa{3}.j^2}{(s_{2j}+s_{3j})(\spb{i}.2 \spa{2}.j + \spb{i}.3 \spa{3}.j)\spa{2}.3}\Bigg)
\cr
&\times\Bigg(\frac{\pi^2}{6} + \frac{1}{2} \ln^2\frac{s_{3j}}{s_{2j}+s_{3j}} + \frac{1}{2} \ln^2\frac{s_{ij}s_{23}}{(s_{2j}+s_{3j})s_{i2}} - \frac{1}{4} \ln^2 \frac{s_{ij} s_{23}}{s_{i2}s_{3j}} \Bigg)
\Bigg\}\,,
\end{align}
\begin{align}
\label{eq:softQuark2}
(\cM^{\rm{u.t.v.}})_{i_2}^{~\ib_3}(2^+,3^-)=-&{ \left(\frac{-s_{2 3}-i\eta}{\mu^2}\right)^{-\epsilon}}
\frac{2}{s_{23}} \frac{13-(20+\delta_R)\epsilon+8\epsilon^2}{2\epsilon(-1+\epsilon)(-3+2\epsilon)(-1+2\epsilon)}
\cr\,
 \times&
 \sum_{i}^mN_c ( t^d )^{~\ib_3}_{i_2}\bf{T}_i^d 
   \frac{ \spa{3}.i \spb{i}.2}{s_{i2}+s_{i3}}\,,
 \cr\,
(\cM^{\rm{s.l.}})_{i_2}^{~\ib_3}(2^+,3^-)=-&{ \left(\frac{-s_{2 3}-i\eta}{\mu^2}\right)^{-\epsilon}}
\frac{2}{s_{23}}\frac{(2\e-1)\e^2\delta_R-2\e^2+3\e-2}{\epsilon^2(-1+2\epsilon)(-1+\epsilon)}
\cr\,
\times&
\frac{1}{2 N_c}
\sum_{i}^m ( t^d )^{~\ib_3}_{i_2} \bf{T}_i^d\frac{ \spa{3}.i \spb{i}.2}{s_{i2}+s_{i3}}\,,
\cr\,
(\cM^{\rm{n_f}})_{i_2}^{~\ib_3}(2^+,3^-)=-&{ \left(\frac{-s_{2 3}-i\eta}{\mu^2}\right)^{-\epsilon}}
\frac{2}{s_{23}}\frac{2-2\epsilon}{\epsilon(-1+2\epsilon)(-3+2\epsilon)}
\cr\,
\times&n_f\sum_{i}^m ( t^d )^{~\ib_3}_{i_2} \bf{T}_i^d 
 \frac{\spa{3}.i \spb{i}.2}{s_{i2}+s_{i3}}\,.
\end{align}
Regarding overall normalizations:
for the single soft gluon   Eqs.~(\ref{eq:treesingle},\ref{eq:ab.})  and double soft quark   Eqs.~(\ref{eq:quark_tree},\ref{eq:softQuark},\ref{eq:softQuark2}),
our expressions carry an overall minus sign compared to Refs~\cite{Catani:1999ss, Catani:2021kcy}.
The minus sign is an overall phase factor and the  cancels in squared amplitude.
Accounting for it, our results   agree with~\cite{ Catani:2021kcy} when external legs are treated in four dimensions.
\section{Simplified differential equation approaches for master integrals}
\label{sec:SDE}
\begin{figure}[h]
  \begin{center}
    \includegraphics[width=0.6\textwidth]{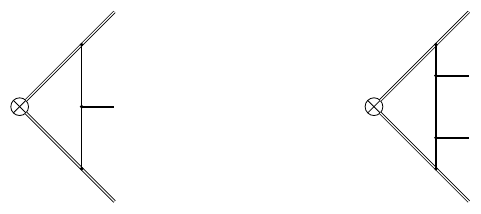}
  \end{center}
  \caption{Topology classifications, with double-line referring to Wilson line}
  \label{fig:7}
\end{figure}
The  master integrals in our problem fall into two topologies,  shown in Fig.~(\ref{fig:7}). The first corresponds to single soft emission.
The propagators  of the  pentagon topology are
\begin{align}
\Bigg\{k^2+i\eta,\,(k+q_2)^2+i\eta,\,(k-q_3)^2+i\eta,\,-n_1\cdot(k+q_2)-i\eta,\,n_4\cdot(k-q_3)-i\eta\Bigg\}\,.
\label{eq:pentagon}
\end{align}
Soft triangles and boxes are  obtained via   standard Feynman parameterization. 
For the soft pentagon integral~\cite{Bern:1993kr}, the traditional differential equations method is involved, 
 here we formulate differential equations with respect to an auxiliary variable~\cite{Papadopoulos:2014lla}. 

To this end, we introduce a family of topologies parameterized by the variable $z$,  by replacing $q_2$ with $z q_2$:
 \begin{align}
\Bigg\{k^2,\,(k+z q_2)^2,\,(k-q_3)^2,\,-n_1\cdot(k+z q_2),\,n_4\cdot(k-q_3)\Bigg\}\,.
\label{eq:topolist}
\end{align}
Each master integral now carries the parameter $z$.
The soft triangles or  boxes follows from replacement $q_2 \to z q_2$ in their closed forms Eq.~(\ref{eq:low point}).
For the  pentagon integral $I_{z;11111}$, we derive a $z$-differential equation treating  the following three rescaling invariants as fixed numbers
\begin{align}
\label{z-change}
x\equiv{s_{1 4}s_{2 3} \over (s_{1 2}+s_{1 3})(s_{2 4 }+s_{3 4})}\,,
\quad
t\equiv\frac{s_{1 2}}{s_{1 2}+s_{1 3}}\,,
\quad
s\equiv \frac{s_{2 4}}{s_{2 4}+s_{3 4}}\,,
\end{align}
and without loss of generality, set
\begin{align}
n_1\cdot n_4=2\,,\quad
q_2\cdot q_3=x/2\,,\quad
n_1\cdot q_2=t\,,\quad
n_1\cdot q_3=1-t\,,\quad
n_4\cdot q_2=s\,,\quad
n_4\cdot q_3=1-s\,.
\end{align}
Then the differential equation with respect to the variable $z$ reads
\begin{align}
\partial_z\bigg( Q(z) I_{z;11111}\bigg)=&
\frac{Q(z)}{z \left(s t z^2-2 s t z+s t+s z-s+t z-t-x z+1\right)}
\cr\,
\times&\bigg\{
\frac{   s t z^2-s t+s+t-1}{(s-1) t z (s z-s+1) (t z-t+1)}\,\epsilon\, I_{z;01100}
    \cr\,
   +&\frac{ s z }{s z-s+1}(2 \epsilon +1)I_{z;11011}
   -s z (2 \epsilon+1)I_{z;11101} 
   \cr\,
   +&\frac{(t-1)}{t z-t+1} (2 \epsilon +1)I_{z;10111} - (t-1) (2 \epsilon +1)I_{z;11110}
   \bigg\}\,,
\end{align}
where $Q(z)$ is an integration factor
\begin{align}
Q(z)\equiv z \left(s t z^2-2 s t z+s t+s z-s+t z-t-x z+1\right)^{\epsilon +1}\,,
\end{align}
 introduced such that the right hand side no longer depends on the pentagon itself.
The integration factor indicates a square root
\begin{align}
Q(z)=&z \left(s t (z - r_1) (z - r_2)\right)^{\epsilon +1}\,,\quad
\Delta(x,s,t)\equiv\sqrt{s^2+4 s t x-2 s t-2 s x+t^2-2 t x+x^2}\,.
\cr\,
r_1=&\frac{-s - t + 2 s t + x - \Delta(x,s,t)}{2 s t}\,,\quad\quad r_2=\frac{-s - t + 2 s t + x + \Delta(x,s,t)}{2 s t}\,.
\end{align}
Regarding boundary conditions: in general the limit $z \to 0$  does not commute with  loop integration,
suggesting an independent    boundary  computation.
However,  since the integration factor is proportional to $z$, it is conjectured  
and  confirmed that $Q(z) I_{z;11111}$ vanishes at the origin provided that $ I_{z;11111}$ is regular there, the same trick was also applied in~\cite{Papadopoulos:2014lla}. 
The Euclidean results for the master integrals  are collected in Eq.~(\ref{eq:low point}) and Eq.~(\ref{eq:five-point}), with a prefactor of $c_\Gamma(-s_{23}-i\eta)^{-\epsilon}$

\begin{align}
I'_{01100}&\equiv\frac{1-2\epsilon}{-2\epsilon}I_{01100}=-\frac{1}{2 \epsilon^2}
\,,
\quad
I_{10011}=\frac{\Gamma(1-2\epsilon)\Gamma(1+\epsilon)}{\epsilon^2\Gamma(1-\epsilon)}
\frac{4}{s_{1 4}}\Bigg(\frac{s_{14}s_{23}}{s_{12}s_{34}} \Bigg)^\epsilon
\,,\cr\,
I_{11011}&=\frac{ \Gamma(1-2\e) \Gamma(1+\e)}{\e^2 \Gamma(1-\e)}\frac{4}{s_{1 2} s_{3 4}}\left(\frac{s_{14}s_{23}}{s_{12}s_{34}}\right)^\e 
\, _2F_1\left(1,1+\e,1-\e,-\frac{s_{2 4} }{s_{3 4}}\right)
\,,
\cr\,
I_{11110}&=-\frac{4}{s_{2 3} (s_{1 2}+s_{1 3})}\frac{1}{\e^2}\,_2F_1\left(1,1,1-\e,\frac{s_{1 3} }{s_{1 2}+s_{1 3}}\right)
\,,
\cr\,
I_{01111}&=\frac{\Gamma(1-2\e)}{\Gamma^2(1-\e)\Gamma(1+\e)}\frac{4}{(s_{1 2}+s_{1 3})(s_{2 4}+s_{3 4})}\left(1-  \frac{s_{1 4} s_{2 3} }{(s_{1 2}+s_{1 3})(s_{2 4}+s_{3 4})} \right)^{-1-\e}
\cr\,
&\times\bigg\{
\frac{2 \Gamma^3(1-\e)\Gamma^2(1+\e)}{\e^2\Gamma(1-2\e)} \left(\frac{s_{1 4} s_{2 3} }{(s_{1 2}+s_{1 3})(s_{2 4}+s_{3 4})}\right)^\e-\frac{2 \Gamma^2(1-\e)\Gamma(1+\e)}{\e^2\Gamma(1-2\e)} 
\cr\,
&\times\, _2F_1\left(-\e,-\e,1-\e,\frac{s_{1 4} s_{2 3} }{(s_{1 2}+s_{1 3})(s_{2 4}+s_{3 4})}\right)
\bigg\}
\,.
\label{eq:low point}
\end{align}
%

%
\begin{align}
&I_{11111}=\frac{4}{s_{12}s_{34}(s_{12}+s_{13})(s_{24}+s_{34})s_{23}}\frac{1}{\epsilon^2}
\cr\,&\times
\Bigg\{s_{34}(s_{12}+s_{13})\Bigg[1+\epsilon\ln\frac{s_{34}(s_{12}+s_{13})}{s_{12}(s_{24}+s_{34})}+\frac{1}{2}\epsilon^2\ln^2\frac{s_{34}(s_{12}+s_{13})}{s_{12}(s_{24}+s_{34})}\Bigg]
\cr\,
&+(s_{12}+s_{13})(s_{24}+s_{34})\Bigg[1-\epsilon\ln\frac{s_{12}s_{34}}{(s_{12}+s_{13})(s_{24}+s_{34})}
+\frac{1}{2}\epsilon^2\ln^2\frac{s_{12}s_{34}}{(s_{12}+s_{13})(s_{24}+s_{34})}\Bigg]
\cr\,
&+s_{12}(s_{24}+s_{34})\Bigg[1+\epsilon\ln\frac{s_{12}(s_{24}+s_{34})}{s_{34}(s_{12}+s_{13})}+\frac{1}{2}\epsilon^2\ln^2\frac{s_{12}(s_{24}+s_{34})}{s_{34}(s_{12}+s_{13})}\Bigg]
\cr\,&-
\frac{s_{14}s_{23}s_{34}(s_{12}+s_{13})}{(s_{12}+s_{13})(s_{24}+s_{34})-s_{23}s_{14}}
\Bigg[\epsilon\ln\frac{s_{23}s_{14}}{(s_{12}+s_{13})(s_{24}+s_{34})}
+\frac{1}{2}\epsilon^2\Bigg(\ln^2\frac{s_{14}s_{23}}{(s_{24}+s_{34})s_{12}}
\cr\,&
+\ln^2\frac{s_{23}s_{14}}{(s_{12}+s_{13})(s_{24}+s_{34})}
-\ln^2\frac{s_{14}s_{23}}{(s_{12}+s_{13})s_{34}}+\ln^2\frac{s_{34}}{s_{24}+s_{34}}-\ln^2\frac{s_{12}}{s_{12}+s_{13}}\Bigg)\Bigg]
\cr\,&-
\frac{s_{14}s_{23}s_{12}(s_{24}+s_{34})}{(s_{12}+s_{13})(s_{24}+s_{34})-s_{23}s_{14}}
\Bigg[\epsilon\ln\frac{s_{23}s_{14}}{(s_{12}+s_{13})(s_{24}+s_{34})}
+\frac{1}{2}\epsilon^2\Bigg(\ln^2\frac{s_{14}s_{23}}{(s_{12}+s_{13})s_{34}}
\cr\,&
+\ln^2\frac{s_{23}s_{14}}{(s_{12}+s_{13})(s_{24}+s_{34})}
-\ln^2\frac{s_{14}s_{23}}{(s_{24}+s_{34})s_{12}}+\ln^2\frac{s_{12}}{s_{12}+s_{13}}-\ln^2\frac{s_{34}}{s_{24}+s_{34}}\Bigg)\Bigg]
\cr\,&+
2 s_{34}(s_{12}+s_{13})\epsilon^2\Bigg[\frac{\pi^2}{6}+\Li_2\left(1-\frac{s_{24}+s_{34}}{s_{34}}\right)\Bigg]
+
2 s_{12}(s_{24}+s_{34})\epsilon^2\Bigg[\frac{\pi^2}{6}+\Li_2\left(1-\frac{s_{12}+s_{13}}{s_{12}}\right)\Bigg]
\cr\,&+
2(s_{12}+s_{13})(s_{24}+s_{34})\epsilon^2\Bigg[\Li_2\left(1- \frac{s_{34}}{s_{24}+s_{34}}\right)+\Li_2\left(1- \frac{s_{12}}{s_{12}+s_{13}}\right)\Bigg]
\cr\,&+
(s_{12}+s_{13})(s_{24}+s_{34})\frac{(s_{12}+s_{13})(s_{24}+s_{34})-s_{23}s_{14}-s_{34}(s_{12}+s_{13})-s_{12}(s_{24}+s_{34})}{(s_{12}+s_{13})(s_{24}+s_{34})-s_{23}s_{14}}
\cr&\times
\epsilon^2\Li_2\left(1-\frac{s_{23}s_{14}}{(s_{12}+s_{13})(s_{24}+s_{34})}\right)-(s_{12}+s_{13})(s_{24}+s_{34})\epsilon^2\frac{\pi^2}{3}
\Bigg\}+\cal O(\e)\,.
\label{eq:five-point}
\end{align}
The following checks are included for the main results of this work,
\begin{itemize}
\item 
By comparing our results with those extracted from the one loop FDH amplitudes $\gamma^* $ $\rightarrow$ $q\bar q gg$ and $\gamma^* $ $\rightarrow$ $q\bar q$ $\bar Q Q$~\cite{Bern:1997sc,Bern:1996ka}, we find agreement.
\item 
On-shell gauge invariance 
\begin{align}
q_{2\mu_2}\bf{J}_{a_2 a_3 }^{\mu_2 \mu_3 (1)}(q_2,q_3) &\propto
\sum_{i=1}^{m} \bf{T}_i^a \;\;,\cr
q_{3\mu_3} \bf{J}_{a_2 a_3 }^{\mu_2 \mu_3 (1)}(q_2,q_3)& \propto
\sum_{i=1}^{m} \bf{T}_i^a \;\;.
\end{align}
\item 
 Numerical checks of the master integrals using toolbox \texttt{pySecDec}~\cite{Borowka:2017idc}.
\end{itemize}

\section{Analytical continuation}\label{sec:AC}
Our results were derived in time-like kinematics with positive Mandelstam variables,
 the corresponding topology with explicit prescription is 
\begin{align}
{{e^+e^-}}:\Bigg\{k^2+i\eta,\,(k+q_2)^2+i\eta,\,(k-q_3)^2+i\eta,\,-n_1\cdot(k+q_2)-i\eta,\,n_4\cdot(k-q_3)-i\eta\Bigg\}\,.
\end{align}
Crossing external partons from the final to the initial state is implemented by reversing the corresponding hard momenta, e.g.
$n_{1}\to -n_{1}$ or $n_{4}\to -n_{4}$.
Equivalently, one may realize   adjust the $i0$-prescriptions of the associated eikonal propagators while keeping the time components positive,
$n_1^0>0,\, n_4^0>0$.
We consider two remaining configurations where we have one or two initial state particles, the corresponding topologies are
\begin{align}
{\rm{SIDIS}}:\Bigg\{k^2+i\eta,\,(k+q_2)^2+i\eta,\,(k-q_3)^2+i\eta,\,-n_1\cdot(k+q_2)+i\eta,\,n_4\cdot(k-q_3)-i\eta\Bigg\}\,,
\cr\,
{\rm{DY}}:\Bigg\{k^2+i\eta,\,(k+q_2)^2+i\eta,\,(k-q_3)^2+i\eta,\,-n_1\cdot(k+q_2)+i\eta,\,n_4\cdot(k-q_3)+i\eta\Bigg\}\,.
\end{align}
In the following, we perform the analytic continuation of the master integrals and associated soft amplitudes from the $e^+e^-$ kinematic region to those relevant for SIDIS and DY topologies.
\subsection{Analytical continuation of the master integrals}
We implement analytic continuation using the following three rescaling invariants
\begin{align}
\label{z-change}
x\equiv{s_{1 4}s_{2 3} \over (s_{1 2}+s_{1 3})(s_{2 4 }+s_{3 4})}\,,
\quad
t\equiv\frac{s_{1 2}}{s_{1 2}+s_{1 3}}\,,
\quad
s\equiv \frac{s_{2 4}}{s_{2 4}+s_{3 4}}\,.
\end{align}
 Reversing a single Wilson line leaves all kinematic invariants unchanged, 
 since the phase factors cancel between numerator and denominator, 
 and no analytic continuation is required.
  In contrast, reversing two Wilson lines generates a non-trivial phase factor associated with the invariant $x$
  \begin{align}
\label{z-change}
x\to x e^{-2\pi i}\,,\quad\quad
t\to t\,,\quad\quad
s\to s\,.
\end{align}
This  indicates a non-trivial monodromy  around the branch   $x=0$, for example
\begin{align}
\lim_{x\to 0}\Li_{2}(1-x)&=-\ln(x)\sum_n \frac{(-x)^n}{n}+\dots=-\ln(x) \ln(1-x)+\dots\,,
\cr
\lim_{x\to 0}\Li_{3}(1-x)&=-\ln(x)\times\frac{-1}{2} \ln^2(1-x)+\dots\,.
\end{align}
where the  ellipsis denotes analytic terms in the neighbourhood of   $x = 0$,
this enables us to compute the discontinuities associated with the crosssing 
\begin{align}
\label{eq:acRules}
{\rm{disc.}}\left[\ln(x)\right]&=-2 \pi i\,,
\cr
{\rm{disc.}}\left[\Li_{2}(1-x)\right]&=2\pi i\, \ln(1-x)\,,
\cr
{\rm{disc.}}\left[\Li_{3}(1-x)\right]&=\pi i \,\ln^2(1-x)\,.
\end{align}
In Section~\ref{sec:SDE} we derived the Euclidean expressions for the master integrals. The analytic continuation to the physical region where both Wilson lines are outgoing (as in configurations such as $\gamma^*\to 4$ partons) is straightforward and follows from
\begin{align}
(-s_{23}-i\eta)^{-\epsilon}\to \abs{s_{23}} e^{i\pi\e}\,.
\end{align}
When some of the Wilson lines correspond to initial-state partons, the continuation rules are given in Eq.~(\ref{eq:acRules}). We conclude:
\begin{itemize}
\item[a)] There is no distinction between the case of two outgoing Wilson lines and the case where one Wilson line is outgoing and the other incoming.
\item[b)] If both Wilson lines are incoming, only the box integral $I_{01111}$ and the pentagon integral $I_{11111}$ develop nontrivial analytic continuation phases.
\end{itemize}
For the soft box $I_{01111}$, the result for two incoming Wilson lines reads
\begin{align}
I^{\rm{ingoing}}_{01111}&=\frac{\Gamma(1-2\e)}{\Gamma^2(1-\e)\Gamma(1+\e)}\frac{4}{(s_{1 2}+s_{1 3})(s_{2 4}+s_{3 4})}\left(1-  \frac{s_{1 4} s_{2 3} }{(s_{1 2}+s_{1 3})(s_{2 4}+s_{3 4})} \right)^{-1-\e}
\cr\,
&\times\bigg\{
\frac{2 \Gamma^3(1-\e)\Gamma^2(1+\e)}{\e^2\Gamma(1-2\e)} \left(\frac{s_{1 4} s_{2 3} }{(s_{1 2}+s_{1 3})(s_{2 4}+s_{3 4})}\right)^\e e^{- 2\pi i\e}-\frac{2 \Gamma^2(1-\e)\Gamma(1+\e)}{\e^2\Gamma(1-2\e)} 
\cr\,
&\times\, _2F_1\left(-\e,-\e,1-\e,\frac{s_{1 4} s_{2 3} }{(s_{1 2}+s_{1 3})(s_{2 4}+s_{3 4})}\right)
\bigg\}
\,.
\end{align}
For the pentagon integral, the corresponding analytic continuation for two incoming Wilson lines is obtained from Eq.~(\ref{eq:five-point}) by implementing the replacements
\begin{align}
\ln\frac{s_{23}s_{14}}{(s_{12}+s_{13})s_{34}}\to&\ln\frac{s_{23}s_{14}}{(s_{12}+s_{13})s_{34}}-2\pi i\,,
\cr\,
\ln\frac{s_{23}s_{14}}{s_{12}(s_{24}+s_{34})}\to&\ln\frac{s_{23}s_{14}}{s_{12}(s_{24}+s_{34})}-2\pi i\,,
\cr\,
\ln\frac{s_{23}s_{14}}{(s_{12}+s_{13})(s_{24}+s_{34})}\to&\ln\frac{s_{23}s_{14}}{(s_{12}+s_{13})(s_{24}+s_{34})}-2\pi i\,,
\cr\,
\Li_2\left(1-\frac{s_{23}s_{14}}{(s_{12}+s_{13})(s_{24}+s_{34})}\right)\to&\Li_2\left(1-\frac{s_{23}s_{14}}{(s_{12}+s_{13})(s_{24}+s_{34})}\right)
\cr\,
+&2\pi i \ln\left(1-\frac{s_{23}s_{14}}{(s_{12}+s_{13})(s_{24}+s_{34})}\right)\,.
\cr\,
\end{align}
\subsection{Universality of the TMD Soft Function in Drell-Yan, SIDIS and $e^+e^-$ Annihilation}
An application of our analytic continuation rules is to provide direct evidence for the statement made in Ref.~\cite{Moult:2018jzp} that the soft functions in Drell-Yan, semi-inclusive deep inelastic scattering (SIDIS), and $e^+e^-$ annihilation are equal up to three loops. Similar observations and discussions can also be found in \cite{Kang:2015moa}. Indeed, we find that
\begin{align}
\cM_{a_2 a_3}^{\rm{SIDIS}}-&\cM_{a_2 a_3}^{\rm{e^+e^-}}=0\,.
\end{align}
\begin{align}
\cM_{a_2 a_3}^{\rm{DY}}(2^+ ,3^+)-&\cM_{a_2 a_3}^{\rm{SIDIS}}(2^+ ,3^+)=
4 { \left(\frac{s_{2 3}}{\mu^2}\right)^{-\epsilon}}\sum_{i \neq j}^2  f_{b c a_2}f_{b d a_3} \bf{T}^c_{i}\bf{T}^d_{j} {\spa i.j \over \spa i.2 \spa 2.3 \spa3.j}
\cr\,
\times&
(- 2 \pi i )
\bigg[
\frac{1}{\e}+\ln \frac{{s_{ij}s_{2 3} \over (s_{i2}+s_{i3})(s_{2j}+s_{3j})}}{1-{s_{ij}s_{2 3} \over (s_{i2}+s_{i3})(s_{2j}+s_{3j})}}
\bigg]+\cal O(\e)\,,
\end{align}
\begin{align}
\cM_{a_2 a_3}^{\rm{DY}}(2^+ ,3^-)-&\cM_{a_2 a_3}^{\rm{SIDIS}}(2^+ ,3^-)=
-4 { \left(\frac{s_{2 3}}{\mu^2}\right)^{-\epsilon}}\sum_{i \neq j}^2  f_{b c a_2}f_{b d a_3} \bf{T}^c_{i}\bf{T}^d_{j} 
\times(- 2 \pi i )
\cr\,
\times&\bigg[
{ {1 \over \spa{i}.2\spb2.j+ \spa i.3 \spb3.j\,}\left({1\over s_{i2}+s_{i3} }{\spb{i}.j \spa i.3^3 \over \spa{2}.3\spa i.2}+{1\over s_{2j}+s_{3j}}{\spa{i}.j\spb2.j^3 \over \spb2.3 \spb3.j}\right)}
\cr\,
\times&\left(
-\frac{1}{\e}+\ln{s_{ij} s_{2 3}\over s_{i2} s_{3j}}
\right)
\cr\,
+&{\spa i.j\over \spa i.2 \spa2.j }{\spb i.j\over \spb i.3 \spb3.j }
\times\bigg(
\ln{s_{ij} s_{2 3}\over s_{i2} s_{3j}}\bigg)
\cr\,
+&{ 1\over s_{i2}+s_{i3}}\Bigg({  -1 \over \spa{i}.2\spb2.j+ \spa i.3 \spb3.j\,}  { \spb{i}.j \spa i.3^3 \over \spa{2}.3\spa i.2}   + {  -1 \over \spb{i}.2\spa2.j+ \spb i.3 \spa3.j\,}  { \spa{i}.j \spb i.2^3 \over \spb{2}.3\spb i.3}  \Bigg) 
\cr\,
 \times&\ln{s_{ij} s_{2 3}\over (s_{i2} +s_{i3} )s_{3j} }
 \cr\,
+&{ 1\over s_{2j}+s_{3j}}\Bigg({  -1 \over \spa{i}.2\spb2.j+ \spa i.3 \spb3.j\,}  { \spa{i}.j \spb2.j^3 \over \spb{2}.3\spb3.j} 
+{  -1 \over \spb{i}.2\spa2.j+ \spb i.3 \spa3.j\,} { \spb{i}.j \spa3.j^3 \over \spa{2}.3\spa2.j}\Bigg)
\cr\,
\times&
\ln{s_{ij} s_{2 3}\over (s_{2j} +s_{3j} )s_{i 2} }
\bigg]
+\cal O(\e)\,.
\end{align}
The difference is proportional to $2\pi i$ and cancel upon adding the complex conjugate contribution. 
Therefore, the virtual-real-real corrections are universal. 
Since the triple-real contribution does not involve any crossing issues, and the existing results for the double-virtual-real~\cite{Duhr:2013msa,Li:2013lsa} and virtual-real-squared~\cite{Catani:2000pi} contributions already display universality due to their overall phase factors, we conclude that the three-loop soft function is universal.
\section{Conclusions}
In this work we presented compact expressions for amplitudes with double soft-gluon and double soft-quark emissions in a generic color configuration. Since gauge redundancy is inherent in the color-space formalism, we found that the amplitudes are most efficiently expressed in helicity variables, which minimize this redundancy~\cite{Dixon:1996wi}. In general, the one-loop soft gluon amplitudes can be decomposed into an Abelian contribution  and a non-Abelian color-correlated term, the latter coupling simultaneously to two hard partons.

We analyzed the strongly ordered limit of the soft amplitudes and derived a factorized representation in terms of two successive eikonal emissions. Furthermore, we investigated the analytic continuation properties of the soft amplitudes and identified non-vanishing discontinuities when crossing into initial-state kinematics. These discontinuities are, however, purely imaginary, implying that the squared amplitude is invariant under crossing. Phenomenologically, this establishes the equivalence of the Drell-Yan, SIDIS, and $e^+e^-$ TMD soft functions up to three loops, a result that had previously been assumed in~\cite{Li:2020bub}.

Combined with the known two-loop single-soft amplitude, which correlates up to three hard partons~\cite{Dixon:2019lnw}, and the tree-level triple-soft emissions that give rise to a quadruple correlation upon squaring~\cite{Catani:2019nqv}, our results clarify the structure of NNNLO soft correlations involving multiple hard directions. This paves the way for extending the calculation of TMD soft functions to configurations with multiple soft Wilson lines and for exploring potential factorization violation effects~\cite{Gao:2019ojf}.

\acknowledgments
I am grateful to Hua Xing Zhu for his continuous support and guidance throughout this project. I also thank Yi Bei Li and Kai Yan for many helpful discussions. The Feynman diagrams were generated using \texttt{feynMF}. This work was supported in part by the National Natural Science Foundation of China (NSFC) under Grant No.~11975200.

Note Added \uppercase\expandafter{\romannumeral1}: 
 I hereby express my gratitude to the authors of Ref.~\cite{Catani:2021kcy} for checking and correcting  part of my original results:
a color factor replacement $-\frac{N_c}{4}+\frac{N_c}{2}\to(-C_F+\frac{N_c}{2})=\frac{1}{2 N_c}$ in the contribution of $\cM^{\rm{s.l.}}$ in Eq.~(\ref{eq:softQuark2}) 
and a phase factor inconsistency (wrong minus sign) in the tree-level expression of the double soft quarks in Eq.~(\ref{eq:quark_tree}).

Note Added \uppercase\expandafter{\romannumeral2}:
I agree with the Erratum listed in the work of~\cite{Czakon:2022dwk}.

\appendix
\section{Gauge invariance and color conservation\label{sec:colorSpace}}
In this section, we introduce color space formalism~\cite{Catani:2000pi} and demonstrate how global  gauge invariance implies color conservation in QCD amplitudes and vice versa.

Consider a generic scattering process that involves $m$ external massless QCD partons and arbitrary number and type of colorless particles in the physical region.
The color space associated with $m$ external partons are tensor product of their individual color representations, a basis vector is
\begin{align}
|c_1,...,c_m\rangle \equiv |c_1\rangle\otimes...\otimes|c_m\rangle\,,
\end{align}
where $c_i$ labels the color index of the $i_{\text{th}}$ parton.
The corresponding scattering amplitude is represented as a vector in this color space 
\begin{align}
&|\cM_m^{s_1,...,s_m;\dots}(p_1,...,p_m)\rangle=\cM_m^{c_1,...,c_m;s_1,...,s_m;\dots}(p_1,...,p_m) |c_1,...,c_m\rangle\,,
\end{align}
 where $s_i$ denotes helicity information and any  additional quantum numbers of parton $i$.
The color charge operator acting on the $i_{\text{th}}$ parton is promoted to an operator in the full color space, 
\begin{align}
T^a_i\to \bf{T}^a_i \equiv\otimes^{i-1}_{1} \mathds{1} \otimes T^a_i\otimes \otimes^{m}_{i+1} \mathds{1} \,,\quad\quad \otimes^{j}_{i} \mathds{1}\equiv\mathds{1}_{i}\otimes\mathds{1}_{i+1}\otimes\dots\otimes\mathds{1}_{j}\,.
\end{align}
The matrix element reads
\begin{align}
\langle c_1,...,c_m|\bf{T}_i^a|b_1,...,b_m\rangle\equiv \delta_{c_1b_1}...T_{c_i,b_i}^a...\delta_{c_mb_m},
 \end{align}
where $T_{cb}^a=i f_{cab}$ if the emitting parton $i$ is a gluon and $T_{\alpha\beta}=t_{\alpha\beta}^a$ 
if the emitting parton is a final state quark or an initial-state antiquark, 
and $T_{\alpha\beta}^a=\bar{t}_{\alpha\beta}^a=-t_{\beta\alpha}^a$ if it is a final-state antiquark or an initial-state quark.
 
To establish the relation between gauge invariance and color conservation, consider the LSZ reduction formula for a generic amplitude,
\begin{align}
\cM^{a,\,i,\,j,\,\dots;s_q,s_{\bar{q}},\dots}&(P_g,P_q,P_{\bar{q}},\dots)=
\nonumber\\
&\bigg\{
\int d X_g e^{-i P_g  X_g}  \varepsilon^{\mu} \square_{\mu\nu} 
\int d X_q e^{-i P_q  X_q} \bar u_{s_q}(P_q)\overrightarrow{ \slash \partial_q}\dots
\nonumber\\
&\langle  \Omega |T\bigg\{A_a^{\nu}(X_g)  \Psi_i (X_q)  \bar \Psi_j (X_{\bar{q}}) \dots \bigg\}|\Omega\rangle
\dots\int d X_{\bar{q}} e^{i P_{\bar{q}}  X_{\bar{q}}}  u_{s_{\bar{q}}}(P_{\bar{q}}) \overleftarrow{\slash  \partial_{\bar{q}}}
\bigg\}_{c.}\,.
\end{align}
Under a global gauge transformation ($\phi=A^\nu,\Psi,\bar\Psi$; $ R= \text{A},\text{F},\text{F}^*$)
\begin{align}
\phi_i\to(1-i\theta ^a T^a_R)_{i\,j} \phi _j\,,
\end{align}
the amplitude must remain invariant, i.e. $|\cM\rangle$ is a color singlet
\begin{align}
e^{-i  \theta \cdot \bf {T}} |\cM\rangle=|\cM\rangle \Longleftrightarrow {\bf T}^a |\cM\rangle=0\,,\quad\quad \bf {T}^a\equiv\sum_i{\bf{ T}^a_i}\,.
\label{eq:gaugeInvariance}
\end{align}
This implies that the total color charge operator annihilates the amplitude,
which expresses color conservation as a direct consequence of global gauge invariance in QCD, representing an application of Noether’s theorem at the amplitude level.

We now turn to the on-shell gauge invariance of the soft current. Under a gauge transformation of the emitted soft gluon, the soft current shifts as
\begin{align}
\varepsilon_\mu(q) \to \varepsilon_\mu(q)+q_\mu\,,\quad\quad \varepsilon(q)\cdot {\bf J}&(q)\to \varepsilon(q)\cdot {\bf J}(q)+ q\cdot {\bf J}(q)\,.
\end{align}
Physical amplitudes must be gauge invariant, so the additional term must vanish when acting on any hard amplitude. Using Eq.~(\ref{eq:gaugeInvariance}), this implies
\begin{align}
q\cdot {\bf J}(q)=&\sum_i{\bf T^a_i}\,,\quad\quad \forall a \in\{1,2,\dots,8\}\,,
\end{align}
which shows that the divergence of the soft current is proportional to the total color charge of the hard subprocess.

\section{Regularization scheme dependence\label{sec:R-dependence}}
The soft amplitude in Eq.~(\ref{eq:plusplus}),\,Eq.~(\ref{eq:plusminus}) and Eq.~(\ref{eq:softQuark}) has dimensional regularization scheme dependence
when computed with $\rm FDH \,(\delta_R=0)$ or $\rm HV \,(\delta_R=1)$  
\begin{align}
\label{eq:diffgluon}
\bf{J}_{a_2 a_3 }^{\mu_2 \mu_3 (1)}(q_2,q_3)|_{\delta_R=0}-\bf{J}_{a_2 a_3 }^{\mu_2 \mu_3 (1)}(q_2,q_3)|_{\delta_R=1}=0+\cO(\e)\,,
\end{align}
\begin{align}
\label{eq:quark}
({\bf J}^{(1)})_{i_2}^{~\ib_3}(q_2,q_3)&|_{\delta_R=0} - ({\bf J}^{(1)})_{i_2}^{~\ib_3}(q_2,q_3)|_{\delta_R=1}={ \left(\frac{-s_{2 3}-i\eta}{\mu^2}\right)^{-\epsilon}}&
\cr
\times&\left(
\frac{1}{2 N_c (\e-1)}-\frac{N_c}{2(\e-1)(2\e-3)(2\e-1)}
\right)
({\bf J}^{(0)})_{i_2}^{~\ib_3}(q_2,q_3)\,,
\cr\,
=&\Bigg(
\frac{1}{6}N_c-\frac{1}{2N_c}
\Bigg)({\bf J}^{(0)})_{i_2}^{~\ib_3}(q_2,q_3)+\mathcal O(\epsilon)\,.
\end{align}
 The difference can be reproduced by considering the soft limit of the one-loop QCD amplitude.
\begin{table}[th]
  \begin{center}
    \footnotesize
    \begin{tabular}{|l|c|c|c|c|}
      \cline{2-4}
      \multicolumn{1}{c|}{}
      & Conventional   & 't~Hooft-- & Four            \\
      \multicolumn{1}{c|}{}
      & dimensional    & Veltman    & dimensional   \\
      \multicolumn{1}{c|}{}
      & regularization &            & helicity           \\
      \cline{2-4}
      \multicolumn{1}{c|}{}
      & CDR            & HV         &FDH                       \\
      \hline
      Number of internal dimensions
      & $d$            & $d$        & $d$             \\
      Number of external dimensions
      & $d$            & 4          & $4$               \\
      Number of internal gluons, $h_g$
      & $d-2$          & $d-2$      & 2                     \\
      Number of external gluons, $n_s(g)$
      & $d-2$          & 2          & 2                      \\
      Number of internal quarks, $h_q$
      & 2              & 2          & 2                      \\
      Number of external quarks, $n_s(q)$
      & 2              & 2          & 2                     \\
      \hline
    \end{tabular}
  \end{center}
  \caption{Definitions of various regularization prescriptions of one-loop
  amplitudes in Ref~\cite{Catani:1996pk}.}
  \label{tab:sumtab}
\end{table}
To this end, we make use of the universal factorization properties of infrared (IR) and ultraviolet (UV) divergences of QCD amplitudes~\cite{Catani:1996pk,Catani:1998bh,Catani:2000ef}, which encode the full dependence on the regularization scheme (RS)~\cite{Gnendiger:2017pys,Gnendiger:2019vnp},
by dimensions of  polarizations (internal or external, see in Table~\ref{tab:sumtab}).

Consider   a \texttt {mass-renormalized} amplitude expanded in the bare strong coupling $g_s$
\begin{equation}
\label{loopex}
\cA_m(g_s, \mu^2; \{p_i,m_i\}) =
\left(\frac{g_s\mu^\epsilon}{4\pi} \right)^q
\,\left[\,\cA_m^{(0)}(\{p_i,m_i\}) 
+ \left(\frac{g_s}{4\pi}\right)^2\,\cA_m^{(1)}(\mu^2; \{p_i,m_i\})
+ {\mathcal O}(g_s^4)\right]\,,
\end{equation}
where $q$ fix the normalization, and $\mu$ is the dimensional-regularization scale. 
In the massless case the one-loop correction $\cA_m^{(1)}$ factorize universally as~\cite{Catani:1996pk,Catani:2000ef,Catani:1998bh}
\begin{equation}
\label{ff1loop}
| \cA_m^{(1)}(\mu^2;\{p_i,m_i\}) \ra_{\RS} =
{\bf I}_m^{\RS}(\epsilon,\mu^2;\{p_i,m_i\})\,
| \cA_m^{(0)}(\{p_i,m_i\}) \ra_{\RS}
+ | \cA_m^{(1)\, {\rm fin}}(\mu^2;\{p_i,m_i\}) \ra
+\mathcal{O}({\epsilon})\:.
\end{equation}
All the $\epsilon$-poles and RS dependence are included in the operator ${\bf I}$, so that 
while the  finite remainder $\cA_m^{(1)\, {\rm fin}}$  is   RS-independent within HV and FDH.
But to remind, the product of the RS-dependent terms
of O$(\epsilon)$ in $\cA_m^{(0)}$ and double poles $1/\epsilon^2$ in ${\bf I}$
produces, in general, a RS dependence of $\cA_m^{(1)}$ that begins at
O$(1/\epsilon)$.
The explicit expression of ${\bf I}_m$ is~\cite{Catani:2000ef}
\begin{align}
\label{iee}
{\bf I}_m^\RS(\epsilon,\mu^2;\{p_i,m_i\}) =&
\frac{(4\pi)^\epsilon}{\Gamma(1-\epsilon)} \Bigg\{
q\,\frac{1}{2}\left( \frac{\beta_0}{\epsilon} - {\tilde{\beta}}_0^{\RS} \right) \cr\,
+& \sum_{j,k=1 \atop k \neq j}^m {\bf T}_j\cdot{\bf T}_k
\left(\frac{\mu^2}{|s_{jk}|} \right)^{\epsilon} \left[
\cV^{(\rm cc)}_{jk}(s_{jk};m_j,m_k;\epsilon)
+ \frac{1}{v_{jk}}\,
\left(\frac{1}{\epsilon}\,i\pi\,
- \frac{\pi^2}{2}\right) \Theta(s_{jk}) \right]  \cr\,
-& \sum_{j=1}^m \Gamma_j^{\RS}(\mu,m_j;\epsilon)
 \Bigg\}\:.
\end{align}
The RS dependence arises from two sources, the first of which is of ultraviolet origin and is proportional to $q$, 
which can be removed by a redefinition of the bare strong coupling $g_s$.
At one loop, it is parametrized by constant coefficient ${\tilde{\beta}}_0^{\RS}$, where
\begin{equation}
\label{betat}
{\tilde{\beta}}_0^{\rm{HV}}=0 \;,\quad\quad\quad {\tilde{\beta}}_0^{\rm{FDH}}= \frac{1}{3} \,C_A\,.
\end{equation}
The second has an infrared origin due to either soft or collinear singularities, and is parametrized by constant coefficient $\Gamma_j^{\RS}$
\begin{align}
\label{cgammag}
\Gamma_g^{\RS}(\mu,m_{\{F\}};\epsilon) =&
\frac{1}{\epsilon} \; \gamma_g - \tilde \gamma_g^{\RS} -
\frac{2}{3}\,T_R \sum_{F=1}^{N_F} \;\ln\frac{m_F^2}{\mu^2}  \;,\cr\,
\Gamma_q^{\RS}(\mu,0;\epsilon) = &\frac{1}{\epsilon} \; \gamma_q - \tilde \gamma_q^{\RS} \,.
\end{align}
The flavour coefficients $\gamma_j$ are
\begin{equation}
\label{cgamma}
\gamma_{j=q,\bar q} = 
\frac{3}{2}\,C_F\:,\qquad\qquad\quad\;\;
\gamma_g = \frac{11}{6}\,C_A - \frac{2}{3}\,T_R \, N_f \:.
\end{equation}
The coefficients $\tilde \gamma_j^{\RS}$ parametrize the finite (for $\epsilon \to 0$)
contributions related to the RS. The transition coefficients $\tilde \gamma_j^{\RS}$ that
relate different RS are given by 
\begin{equation}
\label{cgammatilde}
\tilde{\gamma}^{\rm HV}_j = 0 \:, \qquad
\tilde{\gamma}^{\rm FDH}_{j=q,{\bar q}} = \frac{1}{2} \,C_F \:, \qquad
\tilde{\gamma}^{\rm FDH}_{j=g} = \frac{1}{6} \,C_A.
\end{equation}
With these in hand, we can determine the RS dependence  for the soft amplitude when computed with $\rm{HV}$ or $\rm{FDH}$. According to the factorization theorem
\begin{align}
\cM^{\rm{FDH}}-\cM^{\rm{HV}}\,\,{{\buildrel q_2,q_3 \to 0 \over\longrightarrow}} {\bf J}^{0}    
 \left( |\cM^{(1)}_{\rm{FDH}}\rangle - |\cM^{(1)}_{\rm{HV}}\rangle \right)
 +\left({\bf J}^{(1)}_{\rm{FDH}} - {\bf J}^{(1)}_{\rm{HV}}\right) |\cM^{(0)}\rangle\,.
\end{align}
where ${\bf J}^{0}  $ is the tree-level soft current, ${\bf J}^{(1)}$ is the soft current at one-loop
and $\cM^{(1)}$ is the one-loop QCD virtual correction. For process $\gamma^*$ $\to$ $\qb \,q$, the one-loop virtual corrections read
\begin{align}
\cM^{(1)}_{\rm{FDH}}=&\Bigg(\frac{1}{2} N_c-\frac{1}{2 N_c}\Bigg)\frac{1}{\epsilon^2}(-2-3\epsilon-7\epsilon^2)\cM^{(0)}+\mathcal O(\epsilon)\,,\cr
\cM^{(1)}_{\rm{HV}}=&\Bigg(\frac{1}{2} N_c-\frac{1}{2 N_c}\Bigg)\frac{1}{\epsilon^2}(-2-3\epsilon-8\epsilon^2)\cM^{(0)}+\mathcal O(\epsilon)\,.
\label{virtual}
\end{align}
The difference for double gluon amplitude in QCD is 
\begin{align}
\cM^{\rm{FDH}}&-\cM^{\rm{HV}}\cr\,
&=\Bigg( q \frac{1}{2} \left( (-{\tilde{\beta}}_0^{\rm{FDH}})-(-{\tilde{\beta}}_0^{\rm{HV}})\right) 
+ 2 \tilde{\gamma}^{\rm FDH}_{j=q,{\bar q}}- 2\tilde{\gamma}^{\rm HV}_{j=q,{\bar q}}
+2\tilde{\gamma}^{\rm FDH}_{j=g} -2\tilde{\gamma}^{\rm HV}_{j=g}  \Bigg )\cM^{tree}+\mathcal O(\epsilon)\cr\,
&\,\,{{\buildrel q_2,q_3 \to 0 \over\longrightarrow}}\Bigg( 2 \frac{1}{2} (-\frac{1}{3} C_A) 
+ C_F + \frac{2}{6} C_A \Bigg ){\bf J}^{0}\cM^{(0)}+\mathcal O(\epsilon)\cr\,
&=\Bigg(\frac{1}{2} N_c-\frac{1}{2 N_c}\Bigg){\bf J}^{0} \cM^{(0)}+\mathcal O(\epsilon)\,.
\label{fullgluon}
\end{align}
By inspection, we observe a cancellation between Eq.~(\ref{virtual}) and Eq.~(\ref{fullgluon}), which leads to the result in Eq.~(\ref{eq:diffgluon}). 
For the double soft quark amplitude, we obtain
\begin{align}
\cM^{\rm{FDH}}-\cM^{\rm{HV}}&=\Bigg( q \frac{1}{2} \left( (-{\tilde{\beta}}_0^{\rm{FDH}})-(-{\tilde{\beta}}_0^{\rm{HV}})\right) 
+ 4 \tilde{\gamma}^{\rm FDH}_{j=q,{\bar q}}- 4\tilde{\gamma}^{\rm HV}_{j=q,{\bar q}}  \Bigg )\cM^{(0)}+\mathcal O(\epsilon)\cr\,
&\,\,{{\buildrel q_2,q_3 \to 0 \over\longrightarrow}}\Bigg( 2 \frac{1}{2} (-\frac{1}{3} C_A) +2 C_F  \Bigg ) {\bf J}^{0} \cM^{(0)}+\mathcal O(\epsilon)\cr\,
&=\Bigg(-\frac{1}{3} C_A +2 C_F \Bigg){\bf J}^{0} \cM^{(0)}+\mathcal O(\epsilon)\,.
\label{fullquark}
\end{align}
Subtracting the difference in virtual corrections yields
\begin{align}
{\bf J}^{(1)}_{ \rm{FDH} } - {\bf J}^{(1)}_{ \rm{HV} }
=&\frac{1}{ \cM^{(0)} }\Bigg( (\cM^{ \rm{FDH} }
-\cM^{ \rm{HV} })|_{q_2\to 0,q_3\to 0} - {\bf J}^{0}(\cM^{(1)}_{ \rm{FDH} } - \cM^{(1)}_{ \rm{HV} }  ) \Bigg)\cr\,
=&\Bigg(-\frac{1}{3} C_A + C_F\Bigg){\bf J}^{0}+\mathcal O(\epsilon)\cr\,
=&\Bigg(\frac{1}{6}N_c-\frac{1}{2 N_c}\Bigg){\bf J}^{0}+\mathcal O(\epsilon)\,,
\label{quark-difference}
\end{align}
which is in agreement with Eq.~(\ref{eq:quark}).

\bibliographystyle{JHEP}
\bibliography{DoubleSoft}



\end{document}